\documentclass[sigconf]{acmart}

\usepackage[noend]{algpseudocode}
\usepackage{algorithmicx,algorithm}
\usepackage{url}
\usepackage{multirow}
\usepackage{todonotes}
\usepackage{xcolor}

\usepackage{xspace}

\AtBeginDocument{%
  }

\copyrightyear{2025}
\acmYear{2025}
\setcopyright{cc}
\setcctype{by}
\acmConference[CHI '25]{CHI Conference on Human Factors in Computing Systems}{April 26-May 1, 2025}{Yokohama, Japan}
\acmBooktitle{CHI Conference on Human Factors in Computing Systems (CHI '25), April 26-May 1, 2025, Yokohama, Japan}\acmDOI{10.1145/3706598.3713773}
\acmISBN{979-8-4007-1394-1/25/04}

\newcommand{\mytextcolor}[1]{\textcolor{black}{#1}}

\begin{document}

\title{Classroom Simulacra: Building Contextual Student Generative Agents in Online Education for Learning Behavioral Simulation}

\author{Songlin Xu}
\email{soxu@ucsd.edu}
\affiliation{%
  \institution{University of California, San Diego}
  \city{San Diego}
  \state{California}
  \country{USA}
}

\author{Hao-Ning Wen}
\affiliation{%
  \institution{University of California, San Diego}
  \city{San Diego}
  \state{California}
  \country{USA}
}

\author{Hongyi Pan}
\affiliation{%
  \institution{University of California, San Diego}
  \city{San Diego}
  \state{California}
  \country{USA}
}

\author{Dallas Dominguez}
\affiliation{%
  \institution{University of California, San Diego}
  \city{San Diego}
  \state{California}
  \country{USA}
}

\author{Dongyin Hu}
\affiliation{%
  \institution{University of Pennsylvania}
  \city{Philadelphia}
  \state{Pennsylvania}
  \country{USA}
}

\author{Xinyu Zhang}
\email{xyzhang@ucsd.edu}
\affiliation{%
  \institution{University of California, San Diego}
  \city{San Diego}
  \state{California}
  \country{USA}
}

\renewcommand{\shortauthors}{Xu, et al.}

\begin{abstract}
Student simulation supports educators to improve teaching by interacting with virtual students. However, most existing approaches ignore the modulation effects of course materials because of two challenges: the lack of datasets with granularly annotated course materials, and the limitation of existing simulation models in processing extremely long textual data. 
To solve the challenges, we first run a 6-week education workshop from N = 60 students to collect fine-grained data using a custom built online education system, which logs students' learning behaviors as they interact with lecture materials over time. Second, we propose a transferable iterative reflection (TIR) module that augments both prompting-based and finetuning-based large language models (LLMs) for simulating learning behaviors. Our comprehensive experiments show that TIR enables the LLMs to perform more accurate student simulation than classical deep learning models, even with limited demonstration data. Our TIR approach better captures the granular dynamism of learning performance and inter-student correlations in classrooms, paving the way towards a ``digital twin'' for online education.

\end{abstract}

\ccsdesc[500]{Human-centered computing~Human computer interaction (HCI)}

\keywords{Student Simulation, Generative Agents, Classroom Digital Twin}


\begin{teaserfigure}
  \includegraphics[width=\textwidth]{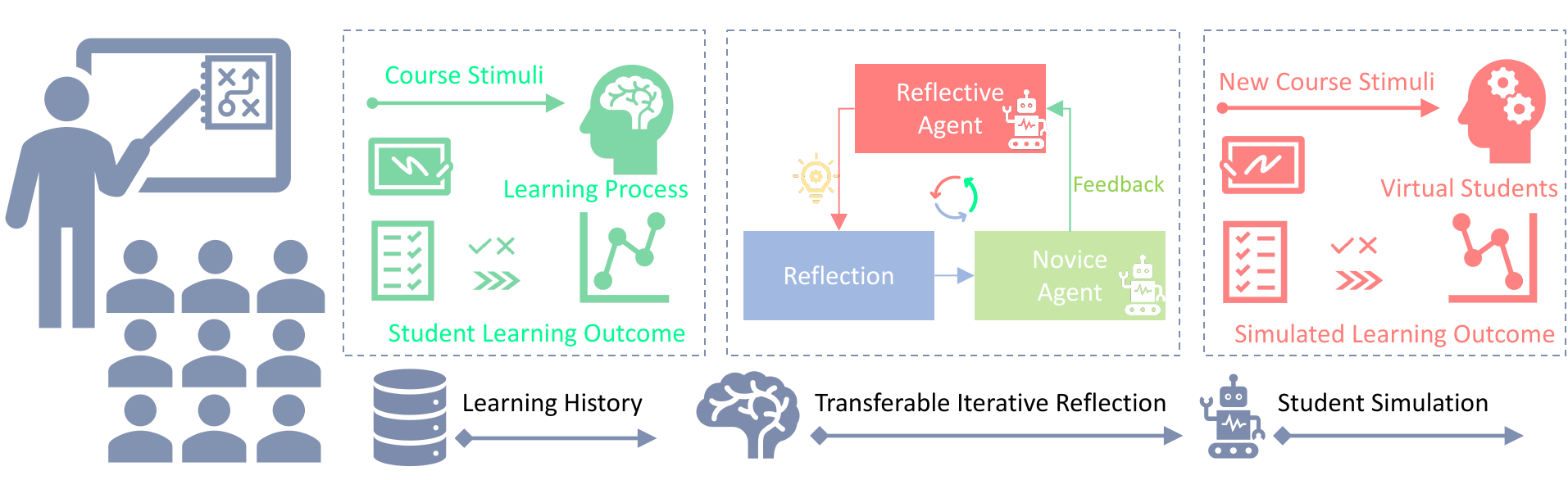}
  \caption{Overview of our Classroom Simulacra framework which uses a transferable iterative reflection method to effectively learn from students' history and capture how course materials modulate learning behaviors, so as to enable realistic student simulation.}
  \Description{
  This figure shows the overview of our Classroom Simulacra framework which uses the transferable iterative reflection to effectively learn from students' history and perform realistic student simulation by capturing the modulation effect of course materials on student learning.
  In learning history, course stimuli are presented to students which incurs a learning process and results in a learning outcome. In transferable iterative reflection, we feed the learning history to the module. Then, we use the iterative interaction between a reflective agent and a novice agent to enable effective reflections from students' learning history and use that reflection to improve student simulation. Finally, in student simulation, new course stimuli are presented to virtual students to simulate the real student's learning outcome with the help of transferable iterative reflection results.}
  \label{teaser}
\end{teaserfigure}

\maketitle

\section{Introduction}
\label{sec:intro}

\mytextcolor{
Accurate simulation of students' learning behaviors in online education settings can help building a ``digital twin'' classroom, which can serve as a high-fidelity sandbox for the instructors to explore diverse pedagogies. This can in turn help improve students' learning performance.} With the rapid advancement of generative AI, using large language models (LLMs) for student simulation is becoming a promising approach. For example, GPTeach \cite{markel2023gpteach} used GPT-based virtual students for interactive TA training and MATHVC \cite{yue2024mathvcllmsimulatedmulticharactervirtual} used LLMs-based virtual classroom for mathematics education. However, these approaches did not systematically evaluate the realism of the virtual students. On the other hand, LLMs-based knowledge tracing models \cite{fu2024sinktstructureawareinductiveknowledge,jung2024clst,li2024integrating,lee2024monacobert,lee2024language} demonstrated high accuracy, but they focus on students' performance prediction rather than multi-faceted behavioral simulation.  
We argue that an accurate digital twin should encompass contextual simulation of students' behaviors, 
capturing the dynamic modulation effect of course materials on both individual students' learning performance and correlations among students. Such dynamism can be reflected in various contextual factors, such as lecture content, individual background, questions, skills, and so on.

However, two main challenges hinder the integration of course materials' effects into student simulation.
The first lies in the lack of fine-grained datasets that annotate course materials along with students' real-time performance. Most existing datasets (such as Ednet \cite{choi2020ednet}, Junyi \cite{JunyiOnlineLearningDataset} and Assistments 2009-2010 benchmark \footnote{https://sites.google.com/site/assistmentsdata/home/assistment-2009-2010-data}) only contain the test question without course materials. The EduAgent dataset \cite{xu2024eduagent} indeed contains the course content, but the lecture length is too short (5 minutes) to reveal the student learning process across a whole lecture. Moreover, students may get tired easily during the course and the resulting data quality may not be guaranteed \cite{wang2019effects}, which unfortunately is not considered by most existing data collection efforts \cite{choi2020ednet,JunyiOnlineLearningDataset}.

A second challenge is that existing language models can only deal with limited contextual data when learning from example demonstrations. LLM-based simulation typically adopts either finetuning-based or prompting-based approaches. The former \cite{lee2024language} enable pre-trained models to learn from new training data directly through model finetuning. But they require a significant amount of computational resources. Therefore, researchers usually resort to smaller language models such as BERT for student simulation \cite{li2024integrating,lee2024monacobert,lee2024language}. However, such smaller language models have very low token limits (e.g., 512 tokens), which can only deal with short textual input but fail to capture complex contextual information such as course materials. Advanced LLMs such as GPT4 could support longer contextual text input, but their performance also drops under a long context \cite{li2024long} and they need significantly more computational resources for fine-tuning.
On the other hand, prompting-based methods \cite{li2024explainable} do not need model training since they directly learn from contextual prompts and example demonstrations. However, the model's in-context learning ability drops significantly in the presence of long demonstrations \cite{li2024long}.

To tackle the first challenge, we run a new user study (N = 60 students and N = 8 instructors for real-time teaching) in the form of a 6-week online education workshop including 12 lectures (1 hour per lecture). We collected student learning performance and fine-grained annotations of course materials and mapped them to specific post-test questions. To guarantee the data quality and improve students' engagement during learning \cite{wang2019effects}, we developed a new online education system that integrated multi-modality sensing techniques to monitor students' cognitive states and prompt instructors to take recommended actions to increase students' learning engagement in real-time.

Furthermore, to deal with the second challenge, inspired by the self-reflection ability of LLMs \cite{madaan2024self} to distill knowledge \cite{Xu2024ASO,gu2024minillm}, we propose a new LLM-based student simulation framework by introducing a transferable iterative reflection (TIR) module, which guides the LLMs to perform reflections on specific course materials and compress the learned knowledge to augment the LLM simulation. Different from a straightforward self-reflection \cite{madaan2024self}, our TIR architecture incorporates iterations between a novice agent and reflective agent to ensure that the reflections could be generalized into new domains to enable transferable reflections (Section \ref{sec:model}). As a result, this TIR module can augment and solve the bottlenecks of both finetuning-based models and prompting-based models. For finetuning-based models, TIR overcomes the token limit issue by focusing its reflections on specific course materials so as to compress the learned knowledge. For prompting-based models, TIR provides an efficient way to enable the LLMs to learn from example demonstrations more effectively, where the learned knowledge can be transferred to 
a new simulation without example demonstrations. This ensures that LLMs learn general knowledge instead of locally optimal knowledge from example demonstrations.  

We have evaluated the student simulation performance in both the EduAgent public dataset \cite{xu2024eduagent} and our newly collected dataset. The results show that our TIR modules enhance the LLM-based student simulation, making it even more powerful than deep learning methods that are trained and fit to given datasets. 
\mytextcolor{
Specifically, the evaluation examines whether our model can better capture the dynamics of student behaviors. Existing research generally defines human behavior as a collection of observable actions and reactions in response to internal (genetic factors) and external (environmental factors) stimuli \cite{longino2019studying}. 
Therefore, to demonstrate that our simulator replicates student learning behaviors, we model student responses to these stimuli, represented by post-test accuracy after engaging with course content. Course knowledge, delivered via lecture slides, constitutes external stimuli, while internal stimuli arise from individual differences such as prior knowledge. Accordingly, we assess the model's ability to capture variations in post-test accuracy at multiple levels: individual (per student), lecture (per session), question (per post-test item), and skill (per knowledge concept). Additionally, we evaluate inter-student correlations to determine whether the simulation model accurately reflects the response patterns between student pairs.
}
Overall, the results show that our approach better captures the granular dynamism of learning performance and inter-student correlations in classroom, pointing towards a potential ``digital twin'' for online classrooms.

To summarize, the main contributions in this paper include:

\begin{itemize}
    \item We run a new 6-week online education experiment with N = 60 students and N = 8 instructors to collect student learning performance with fine-grained annotations of course materials. This is powered by our newly developed online education system that integrates sensing techniques and feedback recommendations to increase student learning engagement during real-time online instruction. The online education system implementation is available at: https://github.com/cogteach-admin/CogTeach, and the data/model implementation is available at: https://github.com/songlinxu/ClassroomSimulacra.
    \item We propose a transferable iterative reflection module that can augment the student simulation performance in both finetuning-based models and prompting-based models.
    \item We systematically evaluate the student simulation performance and show how it can capture the dynamic modulation effect of course materials by revealing lecture-level, individual-level, question-level, skill-level differences and inter-student correlation in classroom.
\end{itemize}

\section{Related Work}
\label{sec:related}


Our work draws inspirations from and advances the knowledge in the following three categories of research.

\subsection{Generative Agents in HCI}


The rapid advancement of LLMs has inspired a wide range of HCI applications, including social behaviors (Generative Agents \cite{park2023generative}), virtual reality \cite{wan2024building} with tour guidance (VirtuWander \cite{wang2024virtuwander}), Human-AI collaboration (AI Chains \cite{wu2022ai}, \cite{wang2024human}), creative tasks (CharacterMeet \cite{qin2024charactermeet}, Luminate \cite{suh2024luminate}, C2Ideas \cite{hou2024c2ideas}, ABScribe \cite{reza2024abscribe}, AngleKindling \cite{petridis2023anglekindling}, PopBlends \cite{wang2023popblends}), healthcare (MindfulDiary \cite{kim2024mindfuldiary}, ChaCha \cite{seo2024chacha}, Narrating Fitness \cite{stromel2024narrating}, \cite{rajashekar2024human}) with health intervention (MindShift \cite{wu2024mindshift}, \cite{jo2024understanding, jo2023understanding, calle2024towards, ma2024evaluating}), web interaction \cite{deng2024large} with UI design (ReactGenie \cite{yang2024reactgenie}, \cite{duan2024generating}, \cite{wang2023enabling}), coding (CollabCoder \cite{gao2024collabcoder}, \cite{liu2023wants}), behavioral change (CatAlyst \cite{arakawa2023catalyst},\cite{bhattacharjee2024understanding}) with human augmentation (Memoro \cite{zulfikar2024memoro}, \cite{jang2024s}), business (Marco \cite{fok2024marco}), and so on.


In educational context, LLMs-powered agents have been utilized to serve as teachable agents (Mathemyths \cite{zhang2024mathemyths}, \cite{jin2024teach}, \cite{liang2023let}) to provide instructions \cite{vadaparty2024cs1}, recommend learning concepts \cite{li2024learning}, and give feedback \cite{matelsky2023large}. For example, DevCoach \cite{wang2024devcoach} supports students' learning in software development at scale with LLMs-powered generative agents. ReadingQizMaker \cite{lu2023readingquizmaker} proposes a Human-NLP collaborative system which supports instructors to design high-quality reading quiz. In programming education, CodeTailor \cite{hou2024codetailor} uses LLM-powered personalized parsons puzzles to support engagement in programming. PaTAT \cite{gebreegziabher2023patat} presents a human-AI collaborative qualitative coding system using LLMs for explainable interactive rule synthesis. 

Such existing work either uses generative agents as the instructors to directly teach students \cite{zhang2024mathemyths} or serves as student agents \cite{markel2023gpteach} to augment intelligent tutoring systems. 
Our work focuses on the second aspect. In what follows, we specifically discuss related work in student simulation using either machine learning or generative agents.

\subsection{Student Simulation}

Student simulation aims to predict student learning behaviors in education, thus providing insights for supporting intelligent tutoring systems \cite{graesser2012intelligent}. A majority of existing research formulates student simulation as a knowledge tracing problem, i.e. predicting students' future learning performance based on their past records  \cite{abdelrahman2023knowledge}. This learning performance is usually represented by the question answering accuracy in the course to measure students' skill levels for specific course concepts. Early work in this domain employed classical Bayesian models \cite{yudelson2013individualized}. In recent years, deep learning models \cite{piech2015deepknowledgetracing} have been the predominant approach for knowledge tracing, combining graph models \cite{nakagawa2019graph}, cognitive theories \cite{zhou2024predictivescalableinterpretableknowledge}, memory-augmented components \cite{zhang2017dynamickeyvaluememorynetworks}, and so on.

\subsection{LLM-based Student Simulation}

Recent work has explored the feasibility of using LLMs directly for predicting students' learning performance \cite{zhang2024predicting,xu2023leveraging} or for knowledge tracing \cite{yu2024eckt} in open-ended questions \cite{liu2022gpt}. 
These methods have better explainability than deep learning models, owing to LLMs' capability to reveal the reasoning process\cite{li2024explainable}.
They can also augment deep learning-based knowledge tracing \cite{fu2024sinktstructureawareinductiveknowledge,jung2024clst}.
In HCI, researchers have developed multi-agent collaboration environment to simulate the whole classroom interaction \cite{zhang2024simulating,chen2023agentverse} and used LLM-simulated student profiles to support question item evaluation \cite{Lu_2024}. These approaches can enable adaptive and personalized exercise generation to augment student learning performance \cite{cui2023adaptive}.
\mytextcolor{For example, Sarshartehrani et al. \cite{sarshartehrani2024enhancing} further leveraged embodied AI tutors for personalized educational adaptation.
GPTeach \cite{markel2023gpteach} also demonstrated the feasibility of using GPT-based virtual students for interactive TA training. In addition, MATHVC \cite{yue2024mathvcllmsimulatedmulticharactervirtual} explored the effectiveness of LLM-simulated multi-character virtual classroom for mathematics education. Moreover, LLMs can help students engage in post-lesson self-reflection \cite{kumar2024supporting} and also support language learning and growth mindset cultivation \cite{kim2024exploring}.}

However, there is also evidence \cite{neshaei2024modelinglearnerperformancelarge} showing the 
limitation of LLMs in student performance prediction compared with deep learning (DL) models. We argue that this is mainly because of the lack of contextual course materials. Specifically, existing LLM-based approaches \cite{fu2024sinktstructureawareinductiveknowledge,jung2024clst,li2024integrating,lee2024monacobert,lee2024language} simply treat student simulation as a sequence prediction problem, which predicts future test performance based on past records, ignoring the modulation effect of course materials. In this case, DL models are very likely to work better than LLMs owing to their capacity to learn from historical data \cite{lecun2015deep, tang2023struc}. In contrast, LLMs are better at few-shot contextual learning based on their large pretrained \textit{knowledge base} \cite{dong2022survey}. Therefore, incorporating contextual course materials could better unleash the power of LLMs to capture the potential effects of course materials on learning performance even with limited data, thus enabling more accurate student simulation.

Existing work in this respect is quite limited, due to both the model limitations and the lack of dataset containing course materials (Section \ref{sec:intro}). 
EduAgent \cite{xu2024eduagent} incorporated course materials to simulate students' cognitive states and post-test performance, but the lectures' duration was too short (5 minutes) to represent the learning process across a typical lecture. By contrast, our work conducts longer-term experiments (6-week, 12 lectures, 1 hour per lecture) to collect high quality learning behavioral data. Our study employs a self-developed online education system integrating sensing techniques and action recommendations to help instructors increase students' learning engagement. Moreover, our proposed transferable iterative reflection module further augments the student simulation performance for both finetuning-based and prompting-based models, which departs from existing approaches in language models \cite{li2024integrating,lee2024monacobert,lee2024language} and deep learning models \cite{piech2015deepknowledgetracing,zhang2017dynamickeyvaluememorynetworks}.

\begin{figure*}
\centering
\includegraphics[width=1\linewidth]{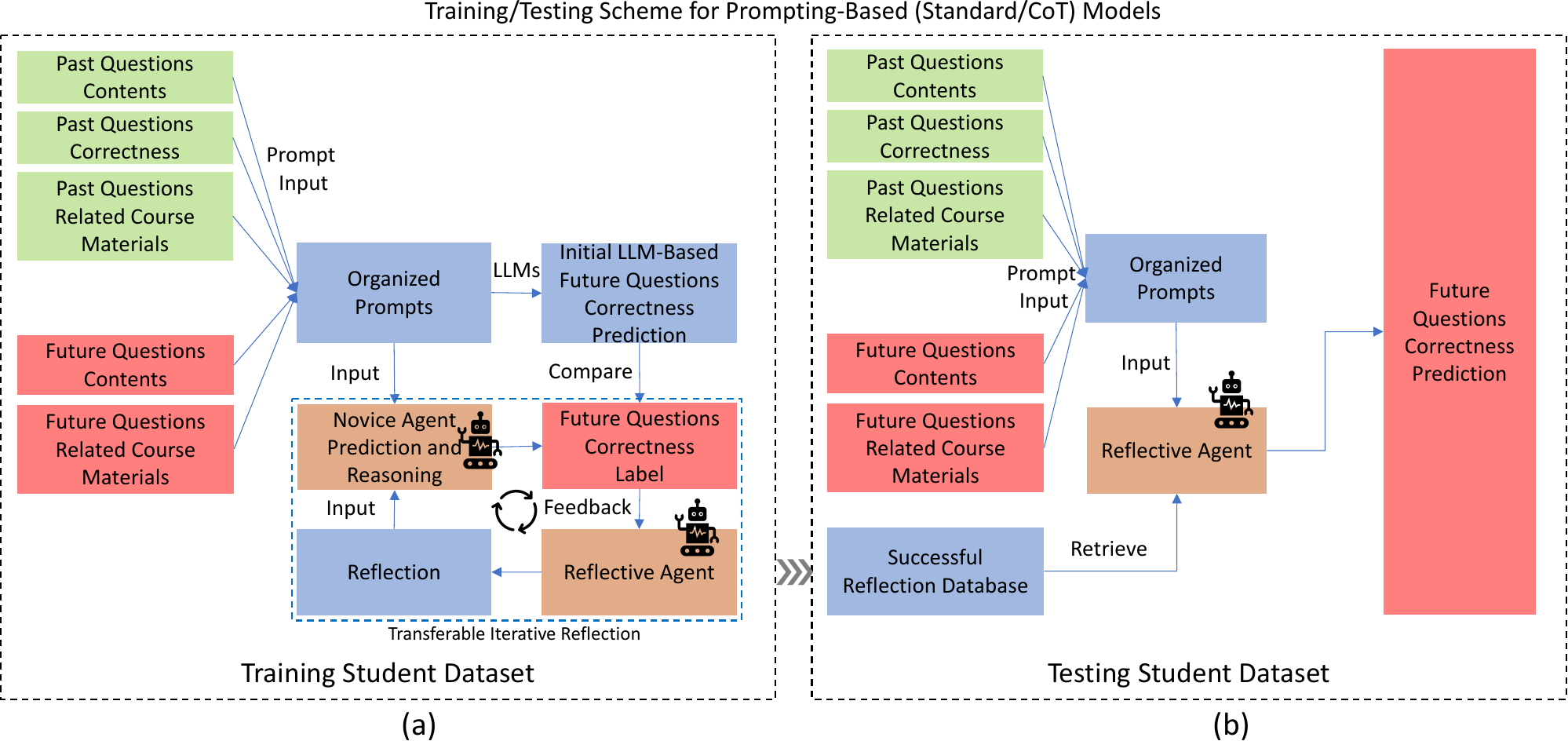}
\caption{\mytextcolor{Training (left) and testing (right) schemes for prompting-based models. }
}
\Description{This figure shows the training (left) and testing (right) schemes for prompting-based models. For both training and testing, we organize past questions contents, correctness, and related course materials along with future questions contents and related course materials as the organized prompt. In training (left figure), we feed the organized prompt to a LLM to generate initial future question correctness prediction. Then, we compare that with the future question correctness label to inform the reflective agent to generate reflections. Finally, we use the generated reflection and organized prompts to make the novice agent make predictions and reasoning. We let the reflective agent and novice agent interact interactively so that the reflective agent has more chances to provide better reflections. In testing (right figure), we use the successful reflection database generated by the reflective agent during training along with the organized prompt to generate future question correctness predictions.}
\label{framework:prompt}
\end{figure*}

\section{Simulation Methodology}
\label{sec:model}
Our classroom simulacra framework aims to build LLM-based generative student agents that could mimic real students' learning behaviors based on their learning histories. The agents can then simulate the students' future learning performance, which is represented by the question answering correctness in the post-lecture tests. 

\subsection{Problem Formulation}
\label{subsec: problem formulate}
\begin{figure*}
\centering
\includegraphics[width=1\linewidth]{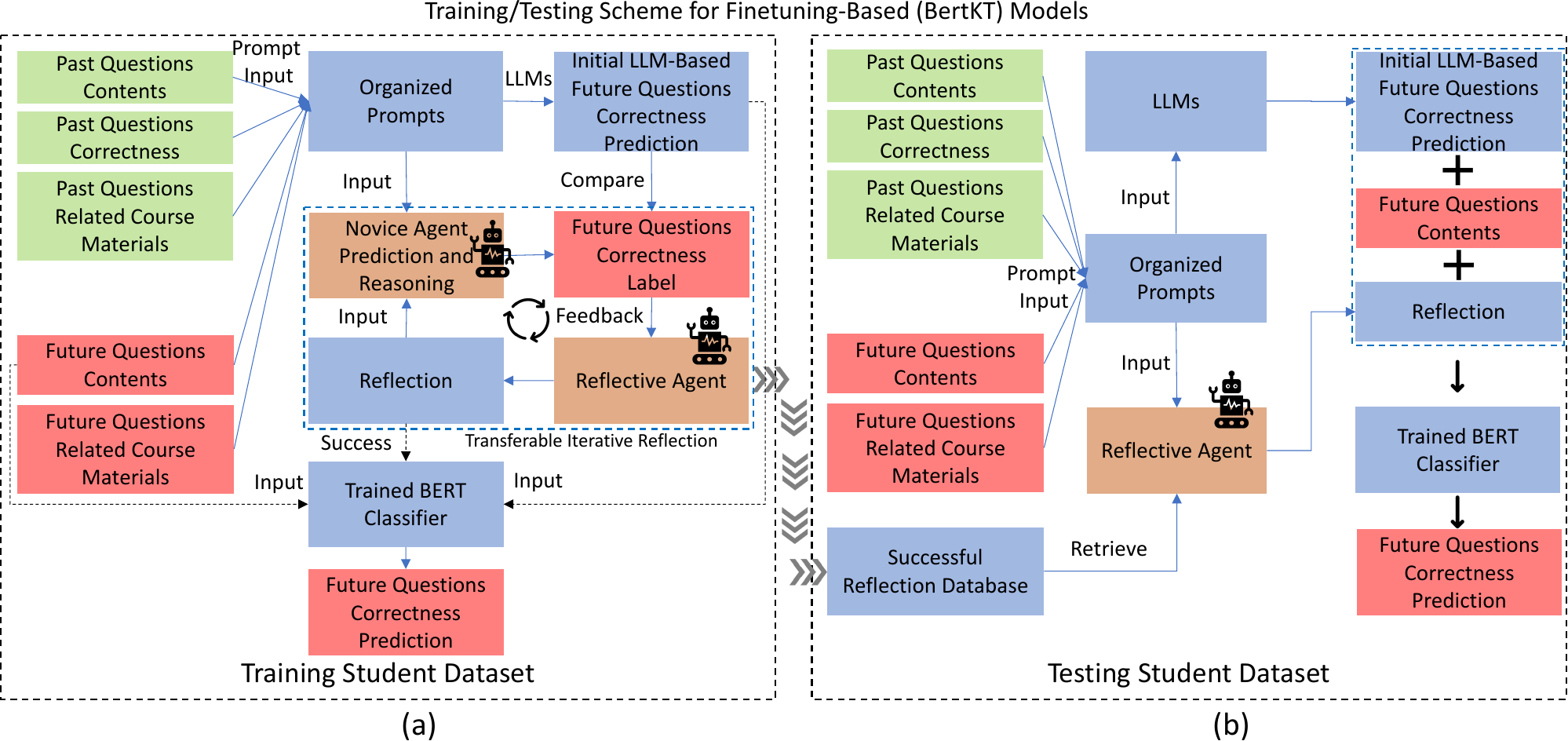}
\caption{\mytextcolor{Training (left) and testing (right) scheme for finetuning-based models. }
}
\Description{This figure shows the training (left) and testing (right) schemes for finetuning-based models. In training, we use a similar TIR approach like prompting-based models to generate reflections. Then, we use the reflection, future question contents, and initial novice agent prediction as input and future questions correctness as labels and output to fine-tune a BERT based classifier.  In testing, we use the successful reflection database generated by the reflective agent during training as the context along with the organized prompt to a new reflective agent to generate the reflection for the future questions. Then, we use the new reflection, new future question contents, and initial novice agent prediction as input to for the trained BERT to predict future question correctness.}
\label{framework:finetune}
\end{figure*}

In our online education scenario, students first listen to the lecture and then finish a post-course test, which evaluates their learning performance based on the accuracy of their answers. As depicted in Fig.~\ref{framework:finetune}, the \textit{input} of our simulation includes past learning history ($l_{past}$) and future learning information ($l_{future}$). Here $l_{past}$ includes past questions' contents ($q_{past}$), past answers' correctness (i.e. labels in past questions, denoted as $y_{past}$), and course materials related to specific questions ($c_{past}$) in the learning history. The future learning information $l_{future}$ includes future question contents ($q_{future}$) and corresponding course materials ($c_{future}$). The \textit{output} of the model is the sequence of future answers' correctness ($\hat{y}_{future}$), with the corresponding ground truth denoted as $y_{future}$.

\mytextcolor{Note that the course material inputs are represented by text only to match the LLM's requirement. Although there might be images in the actual lecture slides, we have converted the images into textual descriptions during our dataset annotation process. 
The course materials include the titles/bullet points on slides, or human-annotated descriptions of images on slides. The lecturers' didactics typically align with the slides but are not delivered as a word-for-word readout. Therefore, we do not use the lecturer's transcripts as course materials nor model input.
Examples of model input can be found in Fig. \ref{prompt example}.}

\subsection{Model Training and Evaluation}
\label{subsec: traing testing split}
We totally have three kinds of simulation models: prompting-based LLM simulation, finetuning-based LLM simulation, and deep learning-based simulation (baseline).
In order to evaluate the models in a comparable manner, all kinds of models use the same training set to train the model and the same testing set to evaluate the simulation performance. However, some prompting-based models only need part of the training set, which will be specified later.
For each dataset, we first split it into training and testing set by following an individual-wise manner with a specific ratio $R$\% (detailed ratio is depicted in each experiment). Specifically, all test performance of $R$\% students was used as the training set and all test performance of another (1-$R$\%) students was testing set. The training set was further divided into model training and model validation set following the same individual-wise manner with the same $R$\% ratio. 
The reason why we use the individual-wise dataset splitting is that our classroom simulacra instantiates each digital student based on the corresponding real student's past learning history and simulates that student's future learning process. 
We set the first five questions as past questions ($q_{past}$) of the student history and other questions as future questions ($q_{future}$) for prediction.

\subsection{Transferable Iterative Reflection (TIR)}
\label{subsec: TIR}

The objective of \textbf{t}ransferable \textbf{i}terative \textbf{r}eflection (TIR) is to improve the LLM-based student simulation by learning from the students' past learning history more effectively. 
The main difference between the TIR and the existing (multi-round) reflection-based methods \cite{madaan2024self,ji2023towards,yan2024mirror,kumar2024supporting,hui2024rot,wang2024taste,li2024think} lies in the transferable feature in the model design. Traditional reflection methods simply ask LLMs to reflect based on the difference between their predictions and labels. 
In contrast, the TIR module iteratively prompts LLMs to reflect on its previous simulations by comparing with labels in the example demonstrations (i.e. past learning history) so that LLMs could generate general reflections results that could be easily 
transferred to novice LLMs which do not have the example demonstrations. This ensures the generalization of such reflected results to be applied into 
new future learning information. As a result, the reflected results can not only directly improve the simple prompting-based simulation by increasing the data utilization efficiency, but also compress the information while avoiding missing important information 
to improve the finetuning-based simulation.
At a higher level, the TIR module consists of four phases: initial prediction, reflection, testing, and iteration 
(Fig. \ref{framework:prompt}). 

\begin{enumerate}
    \item \textit{Initial prediction:} Ask the LLM (reflective agent) to predict future question correctness (i.e., $\hat{y}_{future}$) based on $l_{past}$ and $l_{future}$, and obtain the initial prediction accuracy ($acc_0$) by comparing $\hat{y}_{future}$ with the ground truth ($y_{future}$), as depicted in Section \ref{subsec: problem formulate} and Fig. \ref{framework:finetune}. 
    \item \textit{Reflection:} Provide the reflective agent with the ground truth of future question correctness (i.e. label: $y_{future}$), and ask it to reflect on why it fails to predict some future answers' correctness. The reflection at iteration $k$ is denoted as $r_k$.
    \item \textit{Testing:} Use $r_k$ together with $l_{past}$ and $l_{future}$ to ask another novice agent which has not experienced the ground truth to make a new prediction. Denote the predictions from the novice agent in this iteration $k$ as $\hat{y}_{novice,k}$.
    \item \textit{Iteration:} Obtain the prediction accuracy ($acc_k$) by comparing $\hat{y}_{novice,k}$ with the ground truth ($y_{future}$). If $acc_k$ is lower than the initial prediction accuracy ($acc_0$), we ask the reflective agent to reflect in a different direction in the next iteration. Otherwise, we inform the reflective agent that the accuracy has indeed improved and it could reflect towards the similar direction. \mytextcolor{The iteration ends when the novice agent achieves 100\% accuracy with the help of $r_k$ or we reach the maximum number of iterations.}
\end{enumerate}

Finally, \mytextcolor{we select the reflection that yields the highest prediction accuracy by the novice agent as the best reflection} ($r_{best}=r_{argmax(acc_k)}$) and log it into the reflection database. This database will be used to augment existing prompting-based LLMs by giving example demonstrations or augment finetuning-based models to provide reflections.

\begin{figure*}
\centering
\includegraphics[width=1\linewidth]{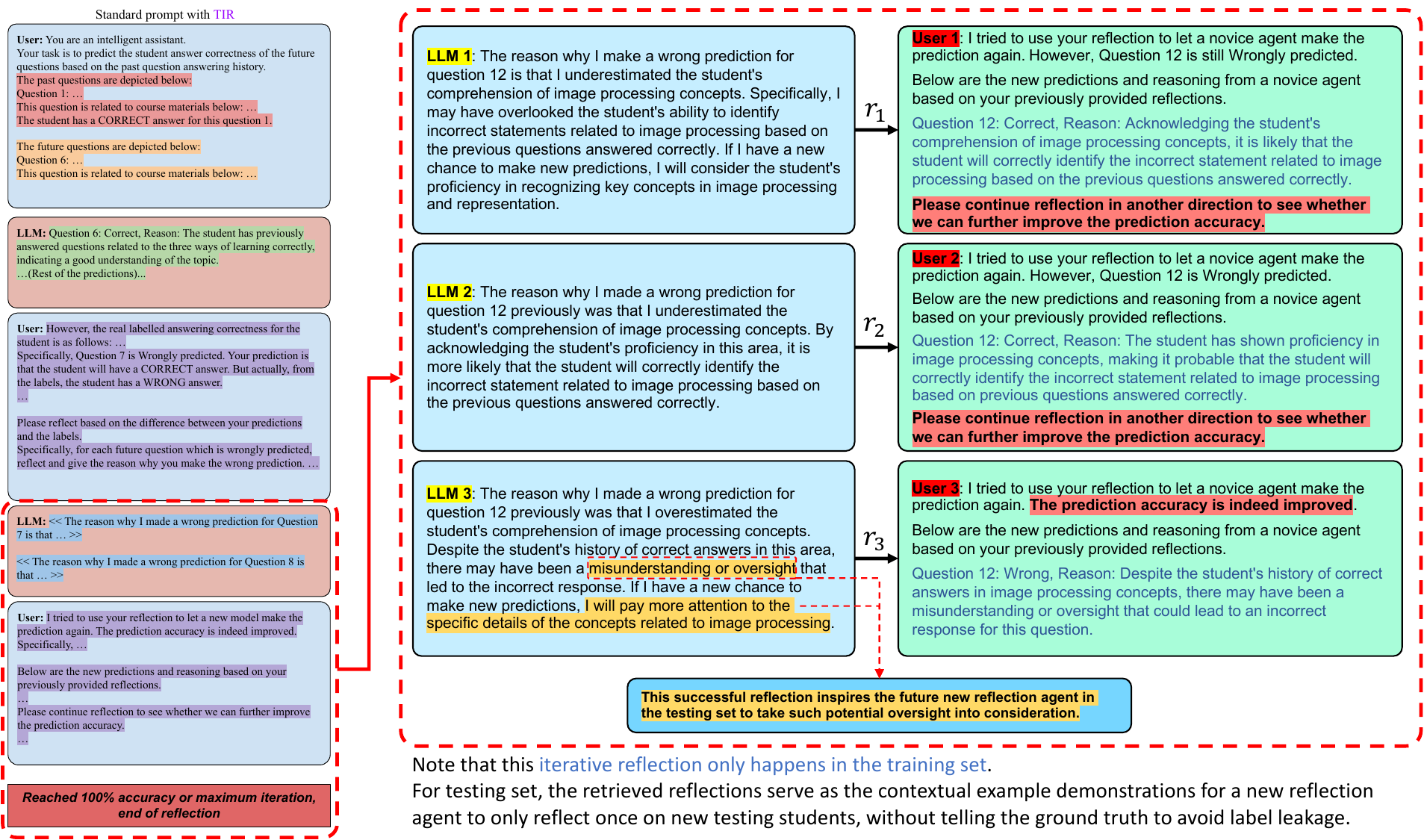}
\caption{\mytextcolor{Prompt examples in the Transferable Iterative Reflection process.}
}
\Description{
This figure shows the prompt examples in the Transferable Iterative Reflection process. 
In the first reflection, LLM says: The reason why I make a wrong prediction for question 12 is that I underestimated the student's comprehension of image processing concepts. Specifically, I may have overlooked the student's ability to identify incorrect statements related to image processing based on the previous questions answered correctly. If I have a new chance to make new predictions, I will consider the student's proficiency in recognizing key concepts in image processing and representation. 
Then user says: I tried to use your reflection to let a novice agent make the prediction again. However, Question 12 is still Wrongly predicted. 
Below are the new predictions and reasoning from a novice agent based on your previously provided reflections.
Question 12: Correct, Reason: Acknowledging the student's comprehension of image processing concepts, it is likely that the student will correctly identify the incorrect statement related to image processing based on the previous questions answered correctly.
Please continue reflection in another direction to see whether we can further improve the prediction accuracy.
In the second reflection, LLM says: The reason why I made a wrong prediction for question 12 previously was that I underestimated the student's comprehension of image processing concepts. By acknowledging the student's proficiency in this area, it is more likely that the student will correctly identify the incorrect statement related to image processing based on the previous questions answered correctly.
Then user says: I tried to use your reflection to let a novice agent make the prediction again. However, Question 12 is Wrongly predicted. Below are the new predictions and reasoning from a novice agent based on your previously provided reflections.
Question 12: Correct, Reason: The student has shown proficiency in image processing concepts, making it probable that the student will correctly identify the incorrect statement related to image processing based on previous questions answered correctly.
Please continue reflection in another direction to see whether we can further improve the prediction accuracy. 
In the third reflection, LLM says: The reason why I made a wrong prediction for question 12 previously was that I overestimated the student's comprehension of image processing concepts. Despite the student's history of correct answers in this area, there may have been a misunderstanding or oversight that led to the incorrect response. If I have a new chance to make new predictions, I will pay more attention to the specific details of the concepts related to image processing. 
Then the user says: I tried to use your reflection to let a novice agent make the prediction again. The prediction accuracy is indeed improved.
Below are the new predictions and reasoning from a novice agent based on your previously provided reflections.Question 12: Wrong, Reason: Despite the student's history of correct answers in image processing concepts, there may have been a misunderstanding or oversight that could lead to an incorrect response for this question.
}
\label{prompt example}
\end{figure*}

\mytextcolor{
The iterative reflection in TIR improves the simulation performance by adjusting the reasoning process of LLMs during reflection. Due to the 
diversity of individuals and course contents, responses to course stimuli varies a lot across students. With a simple reflection, LLMs’ results can be biased due to its pre-training corpus \cite{hadi2023survey}, and cannot be well-adapted to the specific student and course stimuli. The iterative reflection helps LLMs to overcome such bias from its pre-training corpus and iteratively adjust the reasoning process during reflection regarding the causality of how course stimuli modulate students' behaviors while also respecting the students’ past history. 
One example is depicted in Fig. \ref{prompt example}. When the LLM had a straightforward reflection in $r_1$, it overestimated the student’s comprehension and thought the student could answer Question 12 correctly. However, After iterative reflection in $r_3$, the LLM found a potential misunderstanding or oversight that led to the wrong prediction. Such an experience was stored in the successful reflection database. Once retrieved, it could give inspiration for 
the future new reflection agent to consider such oversight in the testing set. 
}

For prompt-based models, TIR can directly improve the LLMs' performance without fine-tuning. For finetuning-based models, TIR can effectively compress the overflowed information in $l_{past}$ and $l_{future}$ to deal with the token limit problem.
More details are described in the following sections.


\subsection{TIR Augments Standard Prompts}
\label{subsec: TIR augments prompts}
Here we describe how to apply the TIR module to augment the standard prompting-based simulation models. 

A standard simulation model simply uses LLMs to take all information as prompt input ($l_{past}$ and $l_{future}$) and predict future question correctness sequence ($\hat{y}_{future}$), as depicted in Fig. \ref{fig:std_prompt_ex}. Directly inputting all of students' data from the training set as example demonstrations poses an obvious challenge. The data including course materials often exceed the token limits of LLMs, hampering their capability to extract useful information.

To this end, we apply the TIR module to enable the LLM to effectively learn from the training data set.
Specifically, in the training stage, we first run the TIR module for each student in the training set, following the procedure in Section \ref{subsec: TIR}. The output reflections are stored into a \textit{successful reflection database}. 
In the testing stage, as depicted in Fig. \ref{framework:prompt}, we do not run the TIR module. Instead, we use a new reflective agent powered by LLMs to retrieve reflections from the successful reflection database. For each simulated student in the testing set, we retrieve the reflections of 
$M$ students from the reflection database. \mytextcolor{The $M$ students from the training set are randomly selected but we make sure they are in the same course as the simulated student in the testing set. This is the only criteria of retrieving reflections, which ensures contextual consistency during reflection. Random selection ensures that the example demonstrations are not manually biased. However, we use a random seed to also ensure that we can replicate such random selection to enhance the replicability of our results. } Our pilot experiment shows that $M=4$ is enough to achieve reasonable simulation performance.
\mytextcolor{
The retrieved reflection from $M$ students serves as the example demonstrations in the same course so that the new reflection agent in the testing set can leverage the experience from the retrieved reflections to perform a transferable reflection.  }
Based on the retrieved reflections and past learning history ($l_{past}$) and future learning information ($l_{future}$), the new reflective agent conducts simulation for the specific student in the testing set.  To prevent label leakage, this new reflective agent does not experience any other training data. So it is different from the reflective agent in the training set.

\begin{table*}[]
\caption{Simulation results on EduAgent dataset}
\Description{This table shows the simulation results on EduAgent dataset.
We found that the integration of the TIR module improved the simulation performance so that both of the simulation accuracy and f1 score were better than all deep learning baseline models. Specifically, the best deep learning model was SimpleKT with 0.6772 accuracy and 0.6698 F1 score. Without the TIR module, the best LLMs-based model was CoT-based prompting with 0.6222 accuracy and 0.5610 F1 score. However, after integrating the TIR module, the best LLMs-based model was finetuning-based BertKT model with 0.7012 accuracy and 0.6880 F1 score, which was superior than the best deep learning model. 
Moreover, we found that the integration of the TIR module could improve all LLMs-based models including standard prompting, CoT prompting, and BertKT. Although the accuracy in CoT slightly decreased, its F1 score was however obviously improved. 
}
\label{tab:result_eduagent}
\begin{tabular}{l|lll|lllll}
\hline
\multirow{2}{*}{Metric} & \multicolumn{3}{l|}{GPT4o-Mini (Without/With TIR)} &  \multicolumn{5}{l}{Deep Learning Models} \\ \cline{2-9} 
                        & Standard  & CoT  & BertKT  & DKT  & AKT  & ATKT  & DKVMN  & SimpleKT  \\ \hline
Accuracy & 0.6025+0.0469  & 0.6222-0.0049  & \textbf{0.6074+0.0938}  & 0.6351  & 0.6171 & 0.6396  & 0.6171  & 0.6772  \\
F1 score     & 0.5128+0.1346  & 0.5610+0.0341  & \textbf{0.6110+0.0770}  & 0.6352  & 0.6051  & 0.6390  & 0.6051  & 0.6698 \\ \hline
\end{tabular}
\end{table*}

\mytextcolor{
It is worth noting that the iterative reflection process only happens in the training set. Moreover, we do not limit the LLMs’ reflection to be either content-specific or metacognitive, in order to give LLMs free enough space to do reflection. But we make it generalizable by evaluating whether the reflections can be transferred to a novice agent for prediction. In addition, the reflection has to be both content-specific and metacognitive, because the course modulation effect is usually different across different lectures. So content-specific reflection is necessary to adapt to specific course context. However, LLMs also have metacognitive reflections because the example demonstration students in the training set are different from the simulated students in the testing set.
In our current setting, \textit{we only need to run the iterative reflection once per lecture to generate the successful reflection database for that speicific lecture}. There is no need to run it for each question/knowledge concept/student. Running reflection offline for each lecture is reasonable, because the lecture materials are usually prepared in advance and available well before the class in real-world teaching scenarios. 
We have not tested if TIR can generalize across different lectures by using one single lecture’s reflection database, but this can be one promising future exploration.
The different reflection direction means that LLMs are instructed to reflect in another reasoning about why a wrong prediction is made. But we do not limit the specific direction content to give LLMs enough space to explore. Examples about such different directions are in Fig. \ref{prompt example} and Appendix Fig. \ref{fig:std_prompt_ex}.  
}

\subsection{TIR Augments CoT-based Models}

Existing work has shown that using the chain of thought (CoT) prompting strategy can improve the capability of LLMs \cite{wei2023chainofthoughtpromptingelicitsreasoning}. The idea of CoT is to use prompts to guide the LLMs to reason step by step like a human, instead of solving the problems all at once.  
Our TIR module can be integrated into CoT to further improve the simulation accuracy of prompting-based models. The integration works similarly to that in standard prompting-based model in Section \ref{subsec: TIR augments prompts}. \mytextcolor{One example workflow is depicted in Appendix Fig. \ref{fig:CoT_prompt_ex}}. The only difference is that we provide step-by-step guidance to the LLMs 
whenever predicting future questions' correctness, as depicted below. 
    
\begin{enumerate}
    \item Analyze the student's past performance:
    \begin{itemize}
        \item Identify course concepts in past questions the student has performed well in and those they have struggled with.
        \item Consider the complexity of the questions and the related course materials.
    \end{itemize}
    \item Review the course concepts and related lecture materials of the future questions:
    \begin{itemize}
        \item Determine the difficulty level of the future questions based on the related course concepts and course materials.
        \item Identify if the future questions are related to the concepts of past questions that the student has previously struggled with or excelled in.
    \end{itemize}
    \item Predict the student's performance in future questions:
    \begin{itemize}
        \item Based on the analysis from steps (1) and (2), predict whether the student will answer each future question correctly or not.
    \end{itemize}
\end{enumerate}

\subsection{TIR Augments Finetuning-based Models}
In addition to prompting-based methods, TIR can also improve finetuning-based language models by compressing the input tokens to avoid token overflow. We fine-tune BERT (Bidirectional Encoder Representations from Transformers), a language representation model that has pre-trained weights \cite{DBLP:journals/corr/abs-1810-04805}. The input of BERT is a sentence and the output could be anything from question answering to semantic classification. However, BERT has very low token limits (512 tokens), which apparently can not directly handle all past/future question inputs or related course materials. The TIR module solves this problem by distilling useful reflections from such data so that there is no need to input the extremely long course materials into BERT.
As depicted in Fig. \ref{framework:finetune}, the input of the TIR augmented BERT is composed of three parts: future question contents $q_{future}$, initial LLMs-based future question correctness prediction results, and reflections from the TIR module. The model output is a binary value to decide whether one student answers one future question correctly or not. This is achieved by finetuning the BERT-based classifier from HuggingFace \footnote{https://huggingface.co/docs/transformers}.
In the training stage, we have the labels for the training set, so the TIR module directly runs on the training set to generate successful reflections, as depicted in Fig.~\ref{framework:finetune}(a) and Section \ref{subsec: TIR}. However, in the testing stage, the labels can not be used for TIR to avoid label leakage. Therefore, we instead use a new \textit{reflective agent} to generate new reflections based on the retrieved reflections as example demonstrations from the successful reflection database in the training set. 

To show the effectiveness of our TIR module in augmenting the BERT model, we prepare another baseline BERT model without TIR, which directly takes all information as input (past questions $q_{past}$ with related course materials $c_{past}$ and real past question correctness labels $y_{past}$, future questions $q_{future}$ with related course materials $c_{future}$) and predict the correctness of future questions.
We denote the BERT model without TIR as \textit{BertKT}, in contrast to that with TIR (\textit{BertKT+TIR}). \mytextcolor{One example of the workflow is depicted in Appendix Fig. \ref{fig:BertKT_io_eg}.}

\mytextcolor{
For a fair comparison, the fine-tuned data is the same as the training set of deep learning models.  Therefore, we only fine-tune the BertKT once in our data. However, for future potential applications to extend the fine-tuned models in external datasets, it is necessary to fine-tune models again in such new datasets, which is similar to deep learning models that use training data to update model weights. 
}

\subsection{Deep Learning Models}

We have also implemented five deep learning models with pyKT \cite{NEURIPS2022_75ca2b23} for student simulation as baseline models, which come from recent state-of-the-art knowledge tracing models. These five models are from four categories: attention-based models (AKT \cite{10.1145/3394486.3403282}, SimpleKT \cite{DBLP:conf/iclr/0001L0H023}), adversarial-based models (ATKT \cite{10.1145/3474085.3475554}), deep sequential models (DKT \cite{piech2015deepknowledgetracing}), and memory-augmented models (DKVMN \cite{zhang2017dynamickeyvaluememorynetworks}). 
\mytextcolor{
These models are widely used knowledge tracing models in the computational education domain to model student learning \cite{10.1145/3394486.3403282, DBLP:conf/iclr/0001L0H023, 10.1145/3474085.3475554, piech2015deepknowledgetracing, zhang2017dynamickeyvaluememorynetworks}. For example, DKT is the first architecture that applies deep learning to model student learning behaviors \cite{piech2015deepknowledgetracing}, which has become the standard baseline model for benchmarking in computational education domain. AKT \cite{10.1145/3394486.3403282} is the first model that applies the monotonic attention mechanism into student modeling.
}
The Appendix Section \ref{appendix: deep learning baseline} includes more details about each model.

The training and testing scheme in deep learning models are the same as other models for fair comparison, as depicted in Section \ref{subsec: traing testing split}. The model input is the same as BertKT, i.e. past questions $q_{past}$ with related course materials $c_{past}$ and real past question correctness labels $y_{past}$, future questions $q_{future}$ with related course materials $c_{future}$. The model output is the prediction of future question correctness $\hat{y}_{future}$. The difference is that deep learning models can not directly take textual data as input. Therefore, we use BERT again to extract embeddings from the textual data (such as question contents and course materials) as deep learning model input, which is a common practice for knowledge tracing models to predict student performance \cite{DBLP:conf/iclr/0001L0H023}.
Each model is initialized using the default configurations in pyKT. The models are trained with Binary Cross Entropy Loss and the AdamW optimizer with a learning rate of 1e-5. Our pilot experiments show that the model validation accuracy stablizes after about 15 epochs. Therefore, we train each model for 30 epochs and select the best model in validation for testing.

\section{Simulation Study}
\label{sec: eduagent results}

We first explored the feasibility of our framework compared with baseline models in the public dataset named EduAgent\cite{xu2024eduagent}. 

\subsection{The EduAgent Dataset}
The EduAgent dataset was collected from N = 301 students, who were asked to watch 5-min online course videos. After that, students were prompted to finish a post test which comprises 10-12 questions. The dataset contained students' correctness on each post test question, as well as corresponding question contents and course materials which were specifically related to each question. More details about this dataset could be obtained from \cite{xu2024eduagent}.

\subsection{Experiment Settings}
We split the dataset into training and testing set by following a individual-wise manner with 0.8 ratio. Specifically, all post test performance of 80\% students were used as the training set and all post test performance of another 20\% students were testing set. The training set was further divided into model training and model validation set following the same individual-wise manner with 0.8 ratio as well. 
We set the first five questions as past questions of the student history and other questions as future questions for prediction.
As depicted in Fig. \ref{framework:prompt}, the simulation model input included the correctness of past questions of real students, as well as corresponding past questions contents and course materials, which were specifically related to each corresponding past question. Moreover, the model input also included future question contents and course materials which are specifically related to each future question. The model output was the correctness of each future question for predictions. 

As depicted in Section~\ref{sec:model}, our TIR module could augment both prompting-based simulation (standard prompt, CoT prompt) and finetuning-based (BertKT) simulation performance. Therefore, in the experiment, we show results of both simulation types with or without the integration of our TIR module. All LLMs-based models used GPT4o-mini. We also compared with five state-of-the-art knowledge tracing models based on deep learning, as depicted in Section. \ref{sec:model}.

\subsection{Results and Analysis}

Results were depicted in Table. \ref{tab:result_eduagent}. We found that the integration of the TIR module improved the simulation performance so that both the simulation accuracy and f1 score were better than all deep learning baseline models. Specifically, the best deep learning model was SimpleKT with 0.6772 accuracy and 0.6698 F1 score. Without the TIR module, the best LLMs-based model was CoT-based prompting with 0.6222 accuracy and 0.5610 F1 score. However, after integrating the TIR module, the best LLMs-based model was finetuning-based BertKT model with 0.7012 accuracy and 0.6880 F1 score, which was superior than the best deep learning model. 

Moreover, we found that the integration of the TIR module could improve all LLMs-based models including standard prompting, CoT prompting, and BertKT, as supported by Table. \ref{tab:result_eduagent}. Although the accuracy in CoT slightly decreased, its F1 score was however obviously improved. 

These results demonstrate the feasibility and effectiveness of our TIR module to enhance existing LLMs-based approaches for more realistic student simulation, which were even better than deep learning models.

\begin{figure}
\centering
\includegraphics[width=1\linewidth]{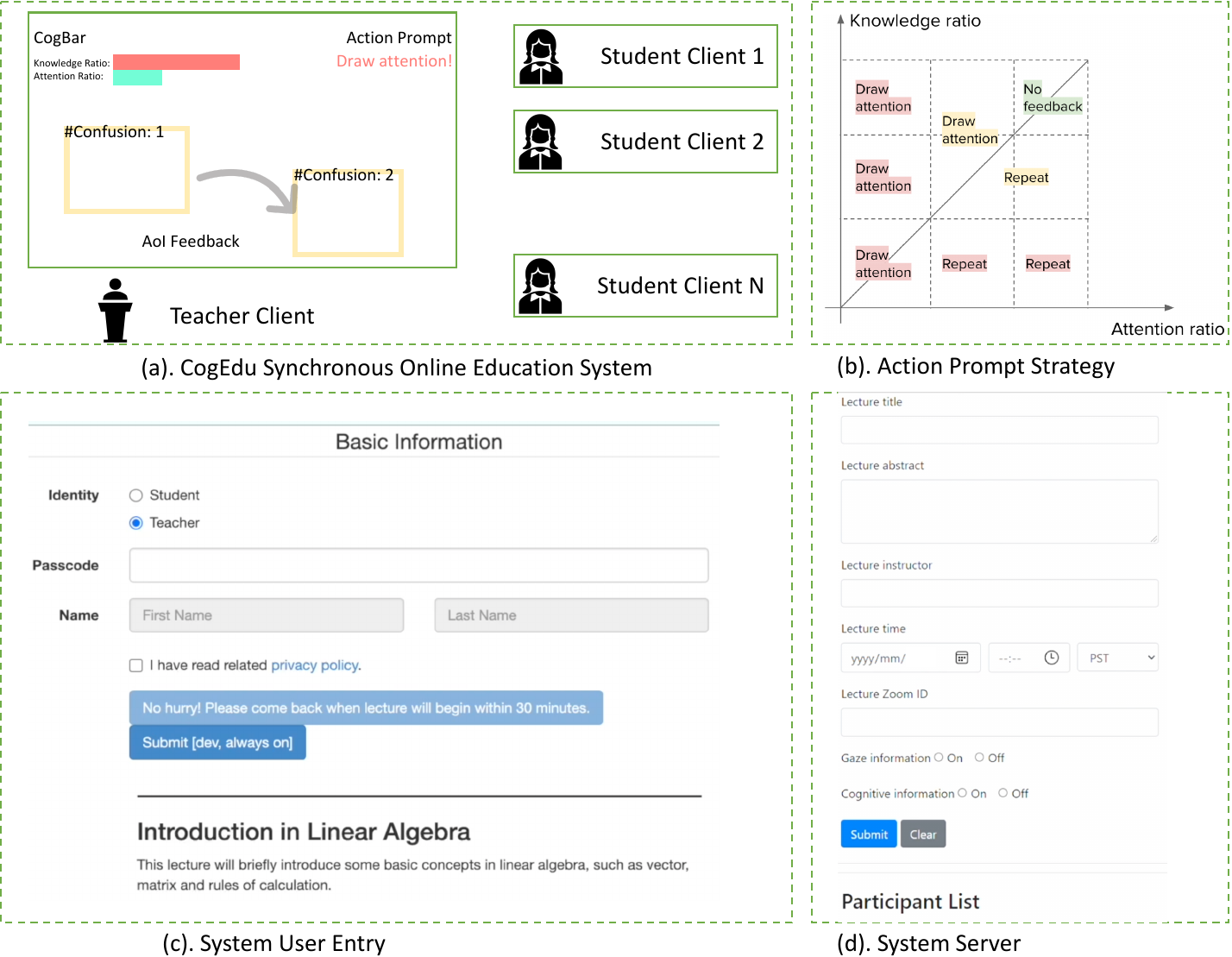}
\caption{
\mytextcolor{
(a). Illustration of our CogEdu system. 
(b). Our action prompt strategy for instructors based on attention ratio and knowledge ratio.
(c,d). The UI of the user end (c) and server end (d) of CogEdu. 
}
}
\Description{Figure (a) shows N student clients on the right side and one teacher client on the left. The teacher client has access to the three modules of feedback: area of interest (AoI) feedback, general cognitive feedback, and action prompt. The bounding boxes of AoI in the middle of the screen represent the area where the students were looking at. The CogBar at the top left corner summarizes the general cognitive feedback. The knowledge ratio, shown as a horizontal bar, is calculated by the ratio of students that were not confused about the contents over all students. The attention ratio, also shown as a horizontal bar below the knowledge ratio, is calculated by the ratio of students that were paying attention over all students. On the top right, the action prompt is the suggestion given to the instructor based on the values of CogBar. In the figure, the action prompt is "Draw attention!" in red. Figure (b) shows our action prompt strategy based on attention ratio and knowledge ratio. Low knowledge ratio triggers "Repeat the current content" recommendation. Low attention ratio triggers "Draw attention" recommendation. The horizontal axis denotes the attention ratio while the vertical axis denotes the knowledge ratio. Each ratio is separated into 3 buckets: low, medium, and high. When both attention and knowledge ratio are high, no feedback was given. When both attention and knowledge ratio are low, "draw attention" was given as the feedback. Other than that, when the knowledge ratio is lower, "repeat" was given as the feedback. When the attention ratio is lower, "draw attention" was given as the feedback to instructors. Figure (c) shows the UI of the user end of CogEdu. There are a few entries for the user to enter: identity (student/teacher), passcode, and name. There is also a lecture title and abstract on the bottom of the page. Figure (d) shows the UI of the server end of CogEdu. There are some entries for the system administrator to enter: lecture title, lecture abstract, lecture instructor, lecture time, lecture Zoom ID, gaze information on/off, and cognitive information on/off. There is also a participant list. 
}
\label{cogedu system}
\end{figure}

\section{Online Education Workshop and Dataset}
\label{sec:workshop}

Although our simulation experiment on the EduAgent dataset demonstrated the effectiveness of our framework compared with baseline models, the EduAgent dataset itself only contains 5-min lectures. Such short duration may not capture the fine-grained effect of course stimuli on student learning performance.
Therefore, it is necessary to examine the student simulation in lectures with longer duration to reveal further insights.

\begin{figure}
\centering
\includegraphics[width=1\linewidth]{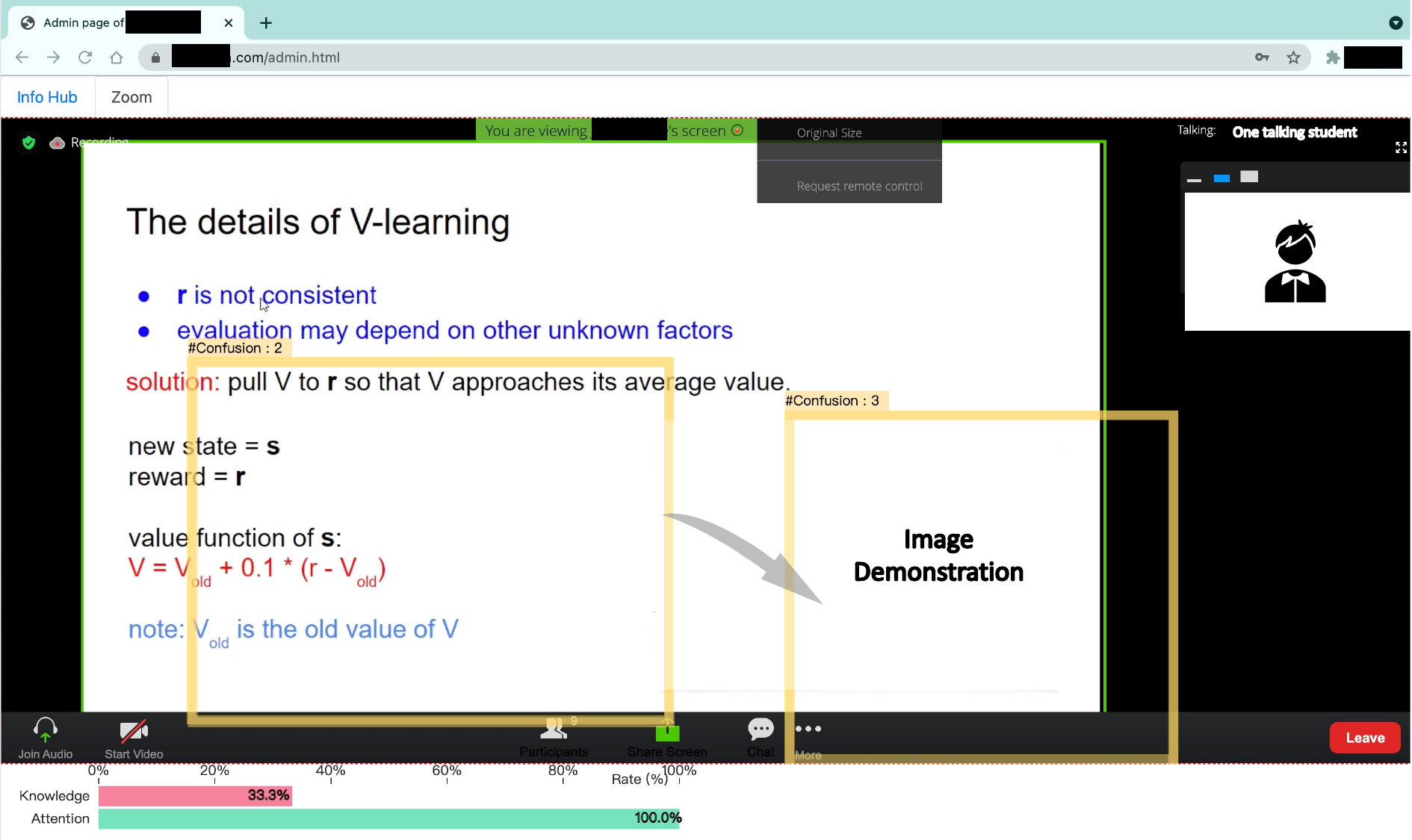}
\caption{A real online education example of our live CogEdu system shown in Fig. \ref{cogedu system}}
\Description{This figure shows a screenshot of a live example of the online education system shown in Fig. \ref{cogedu system}. The system is in a browser that integrates the Zoom interface. In the middle, there is a course slide. In addition, there are two bounding boxes (AoI) on the slide. In the bottom there are two bars: knowledge ratio at 33.3\% and attention ratio at 100\%.}
\label{course system demo}
\end{figure}

\subsection{Workshop Design}

To this end, we conducted a 6-week online education workshop to deliver 12 lectures, where each lecture lasted 1 hour. The long-duration lectures could not only verify the simulation results, but also reveal new insights about how the simulation models can capture students' learning performance variations across the whole lecture (depicted in Section \ref{subsec: dynamism}). \mytextcolor{The workshop syllabus is depicted in Appendix Table. \ref{tab:workshop syllabus}}.

\subsubsection{\textbf{Participants}}

We recruited 30 elementary school students, 30 high school students, and 8 instructors from high schools and universities in the local area using emails and social media. 
We removed the demographic information (such as age and gender) for privacy concerns.
Our data collection was approved by the Institutional Review Board (IRB). All participated students and instructors were informed of the experiment form and then signed consent forms. For participants under 18 years old, we obtained the written consent form from both participants and their parents.

\subsubsection{\textbf{Task and Procedure}}
\begin{figure}
\centering
\includegraphics[width=1\linewidth]{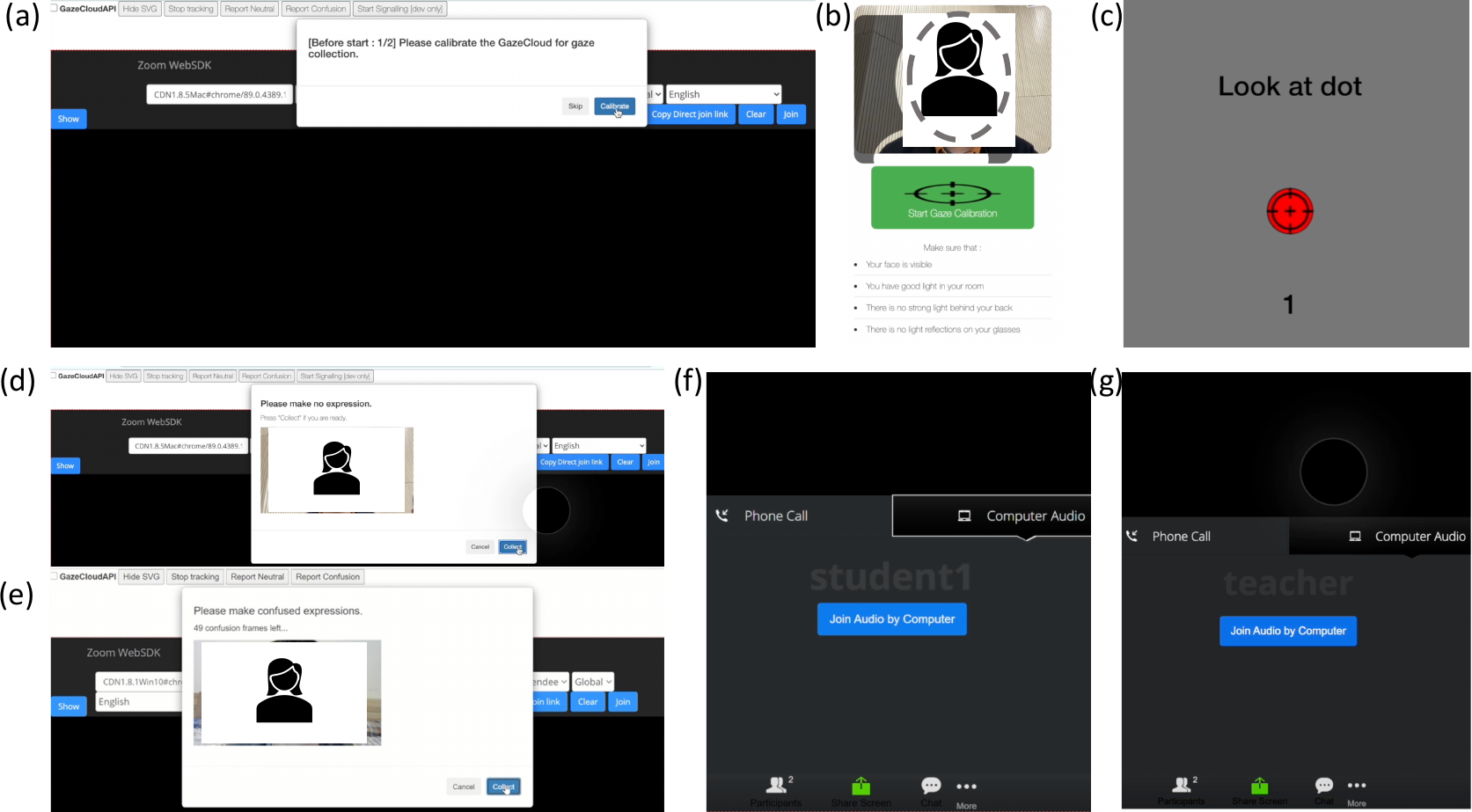}
\caption{
\mytextcolor{The procedure to use our online learning system. (a)(b)(c). Gaze calibration process for gaze tracking. (d)(e). Facial expression model training data collection process. (f)(g). Students and teachers join in the online video calling from their own clients in class.}}
\Description{Figure (a). A popup window saying "[Before start: 1/2] Please calibrate the GazeCloud for gaze collection." (b). A cartoon image of a face on the top. A button saying "Start Gaze Calibration" in the middle. 4 sentences describing what the user must make sure to check at the bottom. (c). A sentence saying "Look at dot". A dot in the middle of the screen. (d). A window saying "Please make no expression." (e). A window saying "Please make confused expression." (f). A window with the word "student" on top of the button "Join Audio by Computer." (g). A window with the word "teacher" on top of the button "Join Audio by Computer."}
\label{system procedure}
\end{figure}

We first prompted the students and instructors to watch an introduction video about how to use our online education system to facilitate learning and teaching, as well as the detailed procedures of our data collection (Fig.~\ref{system procedure}).
After that, students were required to first go through a gaze calibration process (depicted in Section. \ref{subsubsec: student client}) for accurate gaze collection.
Then students were prompted to perform facial expressions (including confused and neutral expressions) in order to train a model for cognitive information detection (more details in Section \ref{subsubsec: student client}). 
After that, both students and instructors were in the same online video conference system (Section \ref{subsec: cogedu}) and instructors presented the course materials to the students. The lecture materials were slides prepared by our research team. During the lectures, our online education system provided visual feedback to the instructors about the students' learning status, and the instructors could adapt their teaching strategies accordingly (depicted in Section. \ref{subsubsec: teacher client}).
After the lecture, students were required to finish a post-test composed of 10-12 questions related to each specific lecture to measure their learning outcome. 

\subsubsection{\textbf{Experiment Design}}
Our six-week workshop was composed of a series of 12 lectures about the basics of artificial intelligence, covering different topics such as basic concepts in machine learning, computer vision, natural language processing, reinforcement learning, etc. Each lecture lasted one hour. The difficulty of the lectures was tailored to match the knowledge level of elementary and high school students, respectively.
For each week, the instructors delivered two lectures. Students were encouraged to select the same time slot for real-time and synchronous teaching among all students together (Fig. \ref{cogedu system}(a)). If students had time conflicts with the instructors, we made new time arrangements for these students for additional data collection.
Each student was encouraged to attend as many lectures as they could. Overall, each student attended 3 lectures on average.

\subsubsection{\textbf{Measurement}}
As depicted above, for students, we mainly collected their gaze, facial expressions, and post-test answers. The gaze and facial expressions were mainly used to generate student status feedback to instructors so that the instructors could take specific actions to increase the students' engagement and improve the quality of collected data. The post-test answers were used to measure the student learning outcomes.

\begin{figure*}
\centering
\includegraphics[width=1\linewidth]{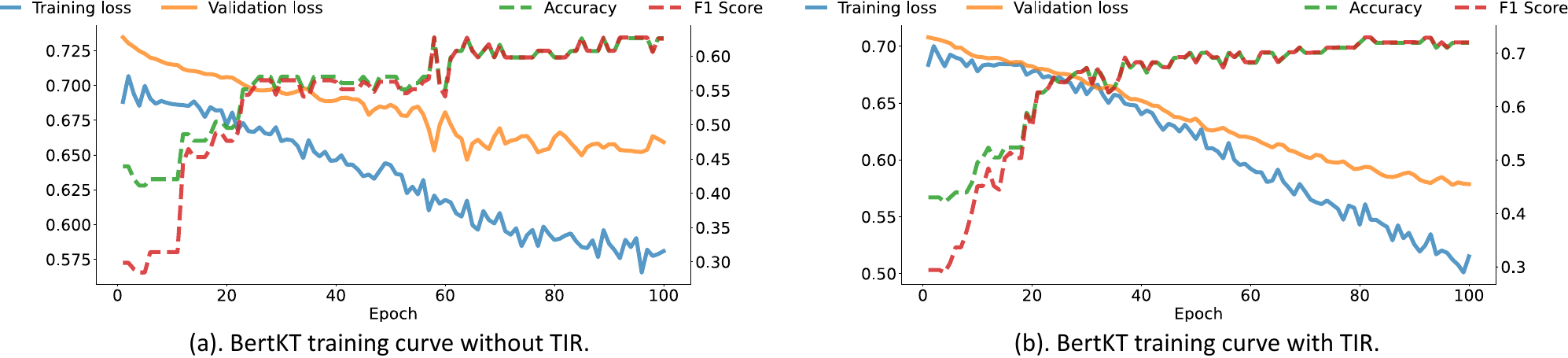}
\caption{
\mytextcolor{
(a)(b). BertKT training curve without (a) and with (b) TIR across different epochs. Vertical axis on the left is the loss value (solid lines). Vertical axis on the right is the metrics (accuracy and F1 score) value (dotted lines).}}
\Description{Figure (a)(b) show the BertKT training curve (training/validation loss, accuracy, and F1 score) with(a) and without(b) TIR across different epochs. Vertical axis on the left is the loss value. Vertical axis on the right is the metrics (accuracy and F1 score) value. The dotted lines represent metrics (accuracy and F1 score) while the solid lines represent training and validation losses. On the top (a), BertKT without TIR training loss curve starts at around 0.69 and ends at around 0.59 with 100 epochs. On the bottom (b), BertKT with TIR training curve starts at around 0.69 and ends at around 0.52 with 100 epochs. On the top (a), BertKT without TIR validation loss curve starts at around 0.73 and ends at around 0.67 with 100 epochs. On the bottom (b), BertKT with TIR validation loss curve starts at around 0.72 and ends at around 0.60 with 100 epochs. On the top (a), BertKT without TIR accuracy curve starts at around 0.45 and ends at around 0.62 with 100 epochs. On the bottom (b), BertKT with TIR accuracy curve starts at around 0.42 and ends at around 0.71 with 100 epochs. F1 scores of BertKT with and without TIR follow similar trends as accuracy scores of BertKT with and without TIR. All the curves are more smooth for BertKT with TIR compared to BertKT without TIR.}
\label{loss}
\end{figure*}

\subsection{Online Education System}
\label{subsec: cogedu}

Although existing video conference software such as Zoom\footnote{https://zoom.us/} provided a stable solution for online education, the subtle student behaviors may not be captured to provide insights to teachers. Moreover, research showed that students' learning performance might become worse compared with in-person instructions\cite{nguyen2015effectiveness}. As a result, the quality of our collected data could be severely compromised. 
Therefore, to solve this problem and facilitate subtle communication between students and instructors, we developed an online education system named \textbf{CogEdu} that could support synchronized teaching between students and teachers in a client on the computer, while also providing real-time student status feedback to teachers to enhance the education process.
Based on the ubiquitous webcams on laptops, we collected the gaze information and facial expressions of students. By analyzing the collected data, we provided real-time feedback to the instructors about the understanding of current contents, the attention status, and a fine-grained visualization of contents that students were concerned about. Understanding about contents (or confusion) and attention were referred to as \textit{cognitive information}. To further assist instructors, the system also provided teaching strategy suggestions based on the collected data. 

As a result, this system could augment student learning engagement and teaching effectiveness to enable high-quality data collection. More details were depicted below.

\subsubsection{\textbf{System Implementation}}

We implemented the CogEdu system on a cloud server. Users (students and instructors) could access the system using their browsers (Fig. \ref{cogedu system}). Considering that most users were more familiar with Zoom, a video teleconferencing software program, we implemented the video conference function based on Zoom APIs\footnote{https://developers.zoom.us/docs/api/}. All the feedback was overlaid over the embedded Zoom interface. To support the large flow of facial expression data before and during the lecture, and to enhance the robustness of the system, we adopted Kubernetes on the google cloud\footnote{https://cloud.google.com/kubernetes-engine} to manage the deployment, scaling, and management. Instances scaled up when the load was growing to reduce latency and achieve satisfying real-time performance.

\subsubsection{\textbf{Student Client}}
\label{subsubsec: student client}
On the student’s side, students were required to first go through a gaze calibration process and then collected facial expressions for cognitive information detection. To collect gaze information from the students, we used the service from GazeRecorder\footnote{https://gazerecorder.com/}. Around 28 gaze positions were provided from the service per second, which were then labeled as fixations or saccades using a velocity-based method \cite{engbert2003microsaccades}. Meanwhile, the system sent facial expressions to the server for cognitive information detection every second. The students' side uploaded all gaze information and cognitive information every five seconds.

The algorithm we used to transform raw gaze into fixations was from \cite{engbert2003microsaccades}. The basic idea was to calculate a velocity threshold, and gaze points with velocity below were labeled as fixations. 
The confusion information of students was detected using a support vector machine (SVM). Before the lecture started, students were asked to make confused expressions and neutral expressions. Collected data were cropped to focus on the eyebrow-eye region and then fed to principal component analysis (PCA) to extract features. An SVM was trained based on the features to classify either confused or neutral expressions.

Attention detection was facilitated by gaze detection, confusion detection, and browser built-in properties. When the user switched to another tab or application, \textit{document.visibilityState} in the browser became hidden. This property was checked together with confusion detection. When the confusion detection algorithm on the server failed to detect a face, which meant the student's face was out of the camera, the system then asserted the student to be not attentive. Thirdly, when the gaze of the student fell out of the screen, the student was labeled as not attentive as well. 

\begin{figure*}
\centering
\includegraphics[width=0.8\linewidth]{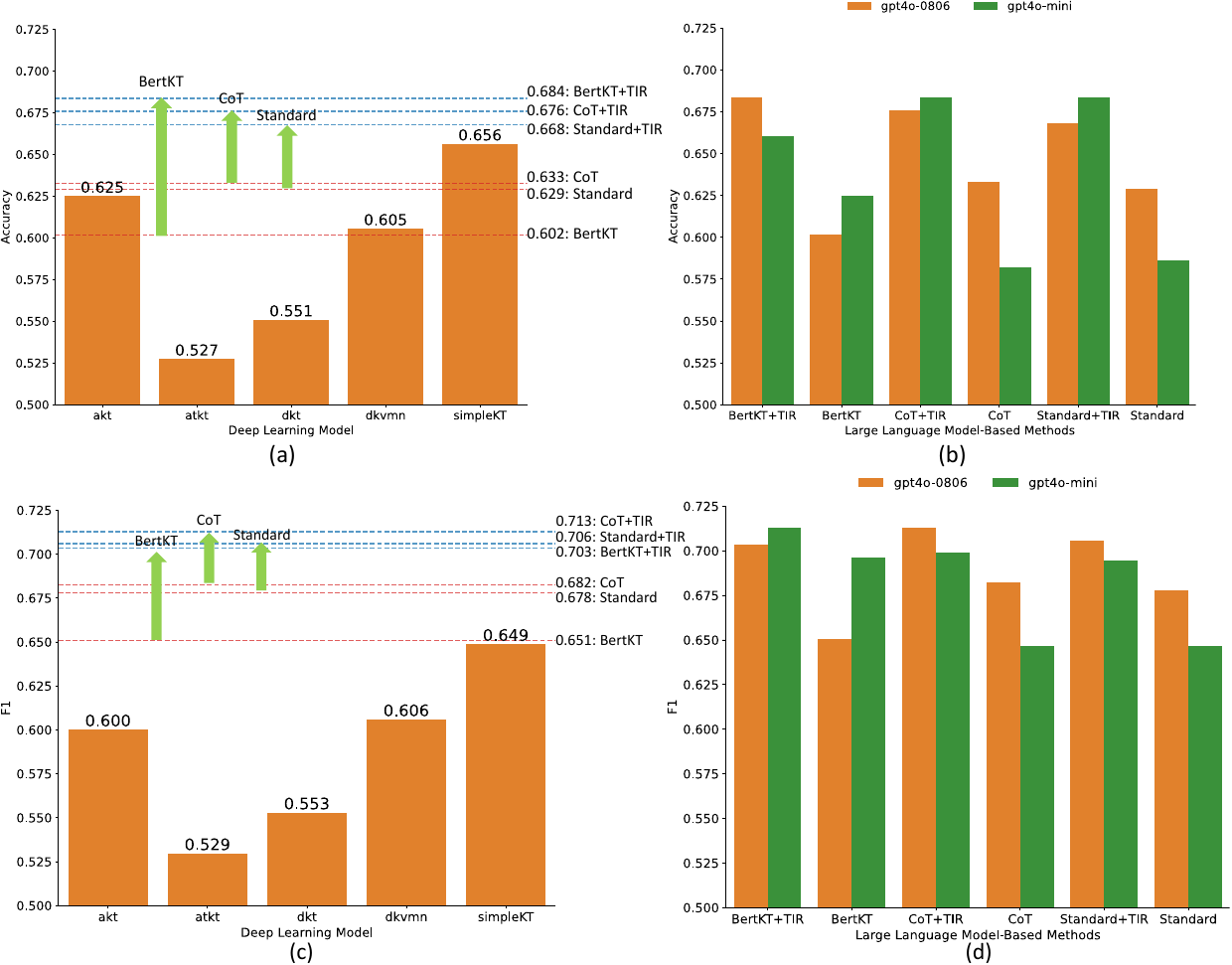}
\caption{
\mytextcolor{
Model accuracy and F1 score comparison on our newly collected dataset. Left (a,c) shows comparison (a: accuracy, c: F1 score) among deep learning models and LLMs-based models using GPT-4o. Right (b,d) shows comparison (b: accuracy, d: F1 score) of LLMs-based models using GPT-4o v.s. GPT-4o mini.}}
\Description{Figure (a) illustrates that large language model based models get improved with TIR and outperform all deep learning baseline models (akt, atkt, dkt, dkvmn, and simpleKT). The accuracy of deep learning models (akt, atkt, dkt, dkvmn, and simpleKT) is 0.625, 0.527, 0.551, 0.606, and 0.656 respectively. The accuracy of large language model based models (Standard, CoT, BertKT) is 0.629, 0.633, and 0.602 respectively. The accuracies of large language model based models with TIR (Standard+TIR, CoT+TIR, BertKT+TIR) are 0.668, 0.676, and 0.684 respectively. Figure (b) illustrates that in most cases (standard and CoT prompts), without TIR, GPT-4o outperforms GPT-4o mini. With TIR, GPT-4o mini outperforms GPT-4o without TIR. Figure (c) illustrates that LLMs-based models get improved with TIR and outperform all deep learning baseline models (akt, atkt, dkt, dkvmn, and simpleKT). The F1 scores of deep learning models (akt, atkt, dkt, dkvmn, and simpleKT) are 0.6, 0.529, 0.553, 0.606, and 0.649 respectively. The F1 scores of LLMs-based models (Standard, CoT, BertKT) are 0.678, 0.682, and 0.651 respectively. The F1 scores of LLMs-based models with TIR (Standard+TIR, CoT+TIR, BertKT+TIR) are 0.706, 0.713, and 0.703 respectively. Figure (d) illustrates the comparison of LLMs-based models with and without TIR using GPT-4o v.s. GPT-4o mini and shows that GPT-4o mini with TIR outperforms GPT-4o without TIR.}
\label{model compare accuracy}
\end{figure*}

\subsubsection{\textbf{Teacher Client}}
\label{subsubsec: teacher client}
On the instructor's side, the system fetched all information that students uploaded to the server every five seconds and then processed the information to provide the instructor with feedback. The feedback provided to instructors consisted of three modules: area of interest (AoI) feedback, general cognitive feedback, and action prompt.

\textbf{Area of Interest Feedback}: This module took as input the gaze and cognitive information and visualized feedback as bounding boxes on the ongoing lecture. These bounding boxes (depicted in Fig. \ref{cogedu system} and Fig. \ref{course system demo}) were clustered from gazes that were labeled as fixations using a spectral clustering algorithm. The bounded area was where students were looking. The opacity of the bounding box represented the ratio of students looking at this area over all students, and the color represented the ratio of students that were confused about the contents in the area over students looking at this area. More details about the spectral clustering method were depicted in Appendix \ref{appendix sub sec: gaze cluster}.

\textbf{General Cognitive Feedback}: This module took as input the cognitive information, and visualized feedback as a summarized bar chart (depicted in Fig. \ref{cogedu system} and Fig. \ref{course system demo}). We displayed the ratio of students that were not confused about the contents over all students (knowledge ratio), and the ratio of students that were paying attention over all students (attention ratio). 

\textbf{Action Prompt}: Based on the general cognitive information, the system provided teaching suggestions to the instructor (depicted in Fig. \ref{cogedu system} and Fig. \ref{course system demo}). When the attention ratio fell lower, instructors were prompted to draw attention from students. When the knowledge ratio fell lower, repeating the current contents was recommended (Fig.~\ref{cogedu system}(b)).


\section{Evaluation}
\label{sec: user study}

Based on the students' data collected from our workshop, we explored the simulation performance in not only straightforward accuracy comparison, but also the dynamic patterns of students' learning performance at fine-grained levels.

\subsection{Simulation Settings}
The simulation settings were similar with those in the EduAgent dataset. 
We split the dataset into training and testing set by following a individual-wise manner with 0.7 ratio. 
We set the first five questions as past questions of the student history and other questions as future questions for prediction.
Both of the model input and output were the same as the EduAgent simulation experiment (Fig. \ref{framework:prompt} and Fig. \ref{framework:finetune}).
For LLMs-based models, we obtained the results with or without our TIR module for both prompting-based models (standard and CoT prompt) and finetuning-based models (BertKT), which were compared with deep learning models. 
All LLMs-based models used both GPT-4o and GPT-4o mini.

\begin{figure*}
\centering
\includegraphics[width=1\linewidth]{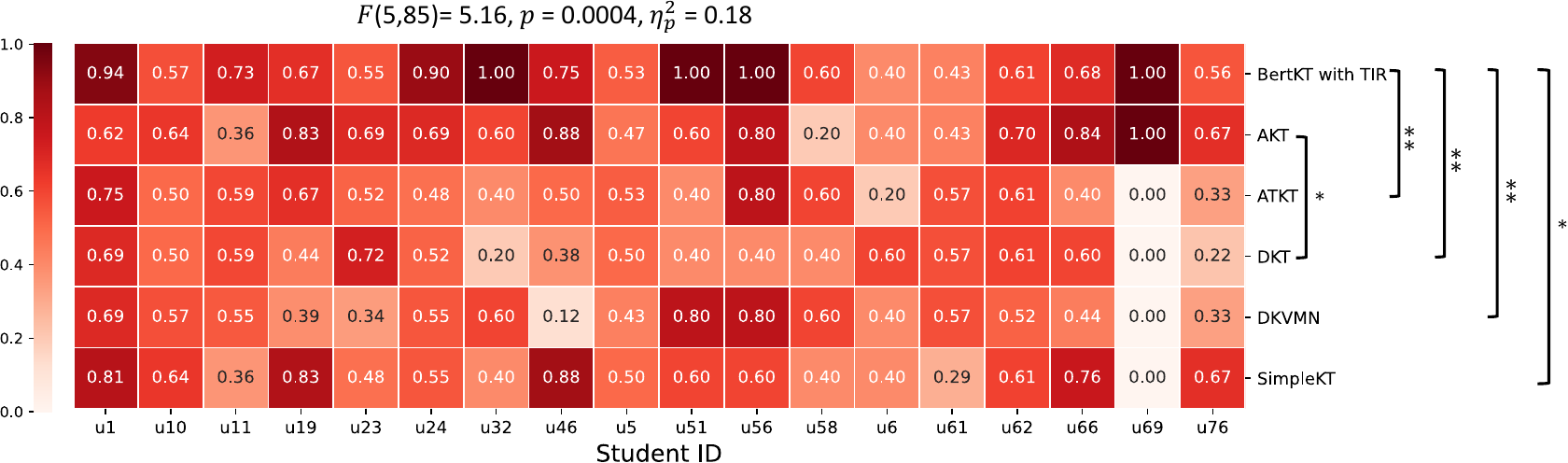}
\caption{
\mytextcolor{
Heatmap to show the average simulation accuracy (each cell) for each individual student using each model.}}
\Description{
This figure shows the average simulation accuracy for each individual student using each model.
We found a significant effect of the model on simulation accuracy ($F(5, 85) = 5.16, \, p = 0.0004, \, \eta_{p}^{2} = 0.18$), indicating a large effect size. For pairwise comparisons, significant differences were observed between the following model pairs:
AKT vs. DKT: $t(17) = 2.29, \, p = 0.035, \, \eta_{p}^{2} = 0.16$.
BertKT with TIR vs. ATKT: $t(17) = 3.42, \, p = 0.003, \, \eta_{p}^{2} = 0.24$.
BertKT with TIR vs. DKT: $t(17) = 3.24, \, p = 0.005, \, \eta_{p}^{2} = 0.30$.
BertKT with TIR vs. DKVMN: $t(17) = 3.83, \, p = 0.001, \, \eta_{p}^{2} = 0.24$.
BertKT with TIR vs. SimpleKT: $t(17) = 2.45, \, p = 0.025, \, \eta_{p}^{2} = 0.14$.
These results indicate that BertKT (with TIR) exhibited significantly better performance compared to most deep learning models. Although we did not find significance between the BertKT (with TIR) and AKT, it still showed better simulation accuracy of BertKT (with TIR) than AKT for most individual students.
}
\label{model compare student}
\end{figure*}

\mytextcolor{
\subsection{TIR Makes LLMs Superior than Deep Learning Models}
We first compared the accuracy and F1-score of the simulated performance of various models by comparing with the real students' performance. 
As depicted in Fig. \ref{model compare accuracy}(a), without the integration of our TIR module, the best model is SimpleKT with 0.656 accuracy, which was better than all LLMs-based models. However, with our TIR module, all LLMs-based models increased the simulation accuracy and all had larger accuracy than the SimpleKT model. The F1 score results were similar as well in Fig. \ref{model compare accuracy}(c), i.e. all TIR-augmented LLMs held better F1 score than all deep learning models. \mytextcolor{These results are encouraging because the selected deep learning models are proven educational models widely accepted in the educational domain \cite{10.1145/3394486.3403282,DBLP:conf/iclr/0001L0H023,10.1145/3474085.3475554,piech2015deepknowledgetracing,zhang2017dynamickeyvaluememorynetworks} to inform teaching practices and support adaptive learning strategies \cite{scholtz2021systematic}.
Therefore, the superior predictive capability of our model indicates a strong potential for real-world applicability.}
To further demonstrate our model's impact, we selected BertKT with TIR as an example to perform statistical analysis by comparing with the five deep learning models in individual-level, lecture-level, and question-level. The effect of different models on student simulation performance was measured by repeated-measures ANOVA with paired t-tests for pair-wise comparisons.
}


\mytextcolor{
\textbf{Individual-Level}: We calculated the average simulation accuracy for each individual student, as depicted in Fig. \ref{model compare student}.
We found a significant effect of the model on simulation accuracy ($F(5, 85) = 5.16, \, p = 0.0004, \, \eta_{p}^{2} = 0.18$), indicating a large effect size. For pairwise comparisons, significant differences were observed between the following model pairs:
\begin{itemize}
    \item AKT vs. DKT: $t(17) = 2.29, \, p = 0.035, \, \eta_{p}^{2} = 0.16$.
    \item BertKT with TIR vs. ATKT: $t(17) = 3.42, \, p = 0.003, \, \eta_{p}^{2} = 0.24$.
    \item BertKT with TIR vs. DKT: $t(17) = 3.24, \, p = 0.005, \, \eta_{p}^{2} = 0.30$.
    \item BertKT with TIR vs. DKVMN: $t(17) = 3.83, \, p = 0.001, \, \eta_{p}^{2} = 0.24$.
    \item BertKT with TIR vs. SimpleKT: $t(17) = 2.45, \, p = 0.025, \, \eta_{p}^{2} = 0.14$.
\end{itemize}
These results indicate that BertKT (with TIR) exhibited significantly better performance compared to most deep learning models. Although we did not find significance between the BertKT (with TIR) and AKT, Fig. \ref{model compare student} still showed better simulation accuracy of BertKT (with TIR) than AKT for most individual students.
}

\begin{figure}
\centering
\includegraphics[width=1\linewidth]{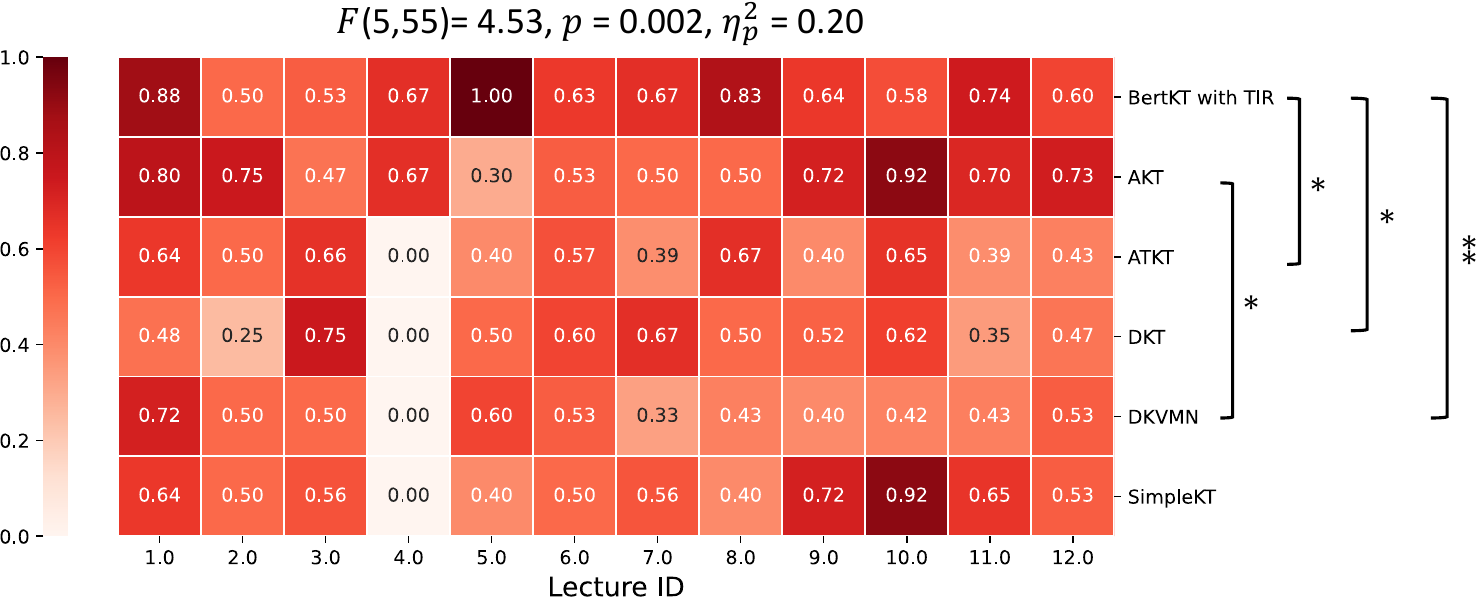}
\caption{
\mytextcolor{Heatmap to show the average simulation accuracy (each cell) for each specific lecture using each model.}}
\Description{
This figure shows the average simulation accuracy for each specific course lecture ID. Results showed a significant effect of the model on simulation accuracy ($F(5, 55) = 4.53, \, p = 0.002, \, \eta_{p}^{2} = 0.20$), indicating a large effect size. Significant differences were observed between the following model pairs:
AKT vs. DKVMN: $t(11) = 2.49, \, p = 0.03, \, \eta_{p}^{2} = 0.21$.
BertKT with TIR vs. ATKT: $t(11) = 3.06, \, p = 0.01, \, \eta_{p}^{2} = 0.28$.
BertKT with TIR vs. DKT: $t(11) = 2.91, \,p = 0.014, \, \eta_{p}^{2} = 0.27$.
BertKT with TIR vs. DKVMN: $t(11) = 4.27, \, p = 0.001, \, \eta_{p}^{2} = 0.35$.
Similar with individual-level results, these results indicate the superiority of the BertKT (with TIR) than these deep learning models. Although we did not find significance between the BertKT (with TIR) and AKT/SimpleKT, it still showed better simulation accuracy of BertKT (with TIR) than AKT/SimpleKT for most lectures.
}
\label{model compare lecture}
\end{figure}

\mytextcolor{
\textbf{Lecture-Level}: We then calculated the average simulation accuracy for each specific course lecture ID, as depicted in Fig. \ref{model compare lecture}. Results showed a significant effect of the model on simulation accuracy ($F(5, 55) = 4.53, \, p = 0.002, \, \eta_{p}^{2} = 0.20$), indicating a large effect size. Significant differences were observed between the following model pairs:
\begin{itemize}
    \item AKT vs. DKVMN: $t(11) = 2.49, \, p = 0.03, \, \eta_{p}^{2} = 0.21$.
    \item BertKT with TIR vs. ATKT: $t(11) = 3.06, \, p = 0.01, \, \eta_{p}^{2} = 0.28$.
    \item BertKT with TIR vs. DKT: $t(11) = 2.91, \,p = 0.014, \, \eta_{p}^{2} = 0.27$.
    \item BertKT with TIR vs. DKVMN: $t(11) = 4.27, \, p = 0.001, \, \eta_{p}^{2} = 0.35$.
\end{itemize} 
Similar with individual-level results, these results indicate the superiority of the BertKT (with TIR) than these deep learning models. Although we did not find significance between the BertKT (with TIR) and AKT/SimpleKT, Fig. \ref{model compare lecture} still showed better simulation accuracy of BertKT (with TIR) than AKT/SimpleKT for most lectures.
}

\begin{figure}
\centering
\includegraphics[width=1\linewidth]{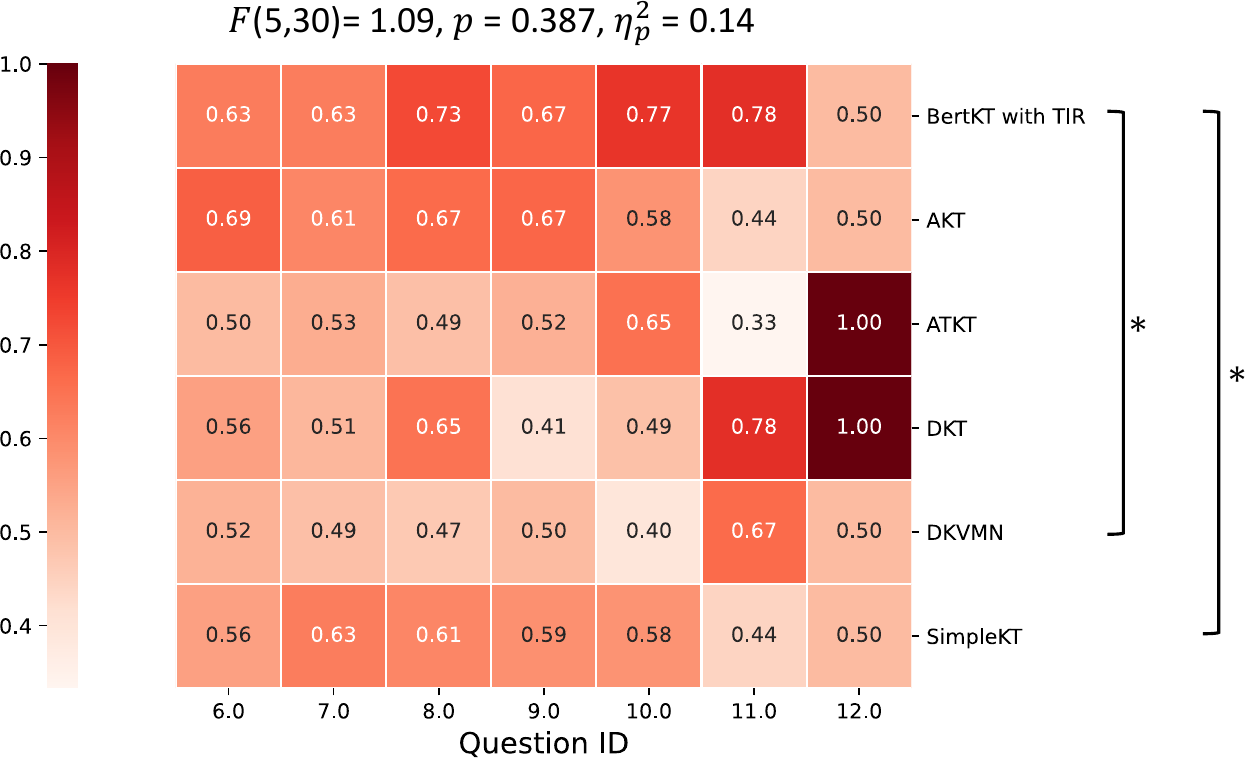}
\caption{\mytextcolor{Heatmap to show the average simulation accuracy (each cell) for each post-test question ID using each model.}}
\Description{
The figure shows the average simulation accuracy for each specific question ID in post-test. 
Results did not find a significant effect of the model on simulation accuracy ($F(5, 30) = 1.09, \, p = 0.387, \, \eta_{p}^{2} = 0.14$). However, significant differences were observed between the following model pairs:
BertKT with TIR vs. DKVMN: $t(6) = 3.69, \, p = 0.01, \, \eta_{p}^{2} = 0.46$.
BertKT with TIR vs. simpleKT: $t(6) = 2.59, \, p = 0.041, \, \eta_{p}^{2} = 0.32$.
Although we did not find significance between the BertKT (with TIR) and AKT/ATKT/DKT, it still showed better simulation accuracy of BertKT (with TIR) than AKT/ATKT/DKT for most question IDs.
}
\label{model compare question}
\end{figure}

\mytextcolor{
\textbf{Question-Level}: We then calculated the average simulation accuracy for each specific question ID in post-test, as depicted in Fig. \ref{model compare question}. 
Results did not find a significant effect of the model on simulation accuracy ($F(5, 30) = 1.09, \, p = 0.387, \, \eta_{p}^{2} = 0.14$). However, significant differences were observed between the following model pairs:
\begin{itemize}
    \item BertKT with TIR vs. DKVMN: $t(6) = 3.69, \, p = 0.01, \, \eta_{p}^{2} = 0.46$.
    \item BertKT with TIR vs. simpleKT: $t(6) = 2.59, \, p = 0.041, \, \eta_{p}^{2} = 0.32$.
\end{itemize}
Although we did not find significance between the BertKT (with TIR) and AKT/ATKT/DKT, Fig. \ref{model compare question} still showed better simulation accuracy of BertKT (with TIR) than AKT/ATKT/DKT for most question IDs.
}

\subsection{TIR Enhances Model Learning Efficiency}
Although the training of BertKT+TIR model used all training students, it is worth noting that, for other prompt-based models (Standard and CoT), we only used four example students as the contextual example demonstration instead of all students in the training set. 
However, after integrating our TIR module, both of prompt-based models (Standard and CoT) achieved better simulation performance than deep learning models (Fig. \ref{model compare accuracy}), which used all students in the training set for model training. This demonstrates that our TIR module could enhance the exploitation efficiency of prompt-based models to achieve comparable or even more realistic student simulation within more limited training data.

\subsection{TIR Empowers Smaller LLMs}
We also compared the simulation performance using both GPT-4o and GPT-4o mini.
We found that the integration of our TIR module increased all of the GPT-4o mini based simulation models (Standard, CoT, BertKT), which were even better than the simulation models using GPT-4o without the TIR module. Note that GPT-4o mini is a much smaller model than GPT-4o\footnote{https://openai.com/index/gpt-4o-mini-advancing-cost-efficient-intelligence/}. Without TIR, GPT-4o had apparently better simulation performance than GPT-4o mini in Standard and CoT models, as depicted in Fig. \ref{model compare accuracy}(b). However, after integrating the TIR module, both Standard and CoT models in GPT-4o mini outperformed GPT-4o in an obvious margin (Fig. \ref{model compare accuracy}(b)). These results demonstrate that the TIR module could improve the smaller LLMs to learn from example students in the training set more effectively. As a result, smaller LLMs could achieve comparable or even better performance, eliminating the need of using larger size LLMs.

\subsection{TIR Captures Individual Differences Better}
\begin{figure*}
\centering
\includegraphics[width=1\linewidth]{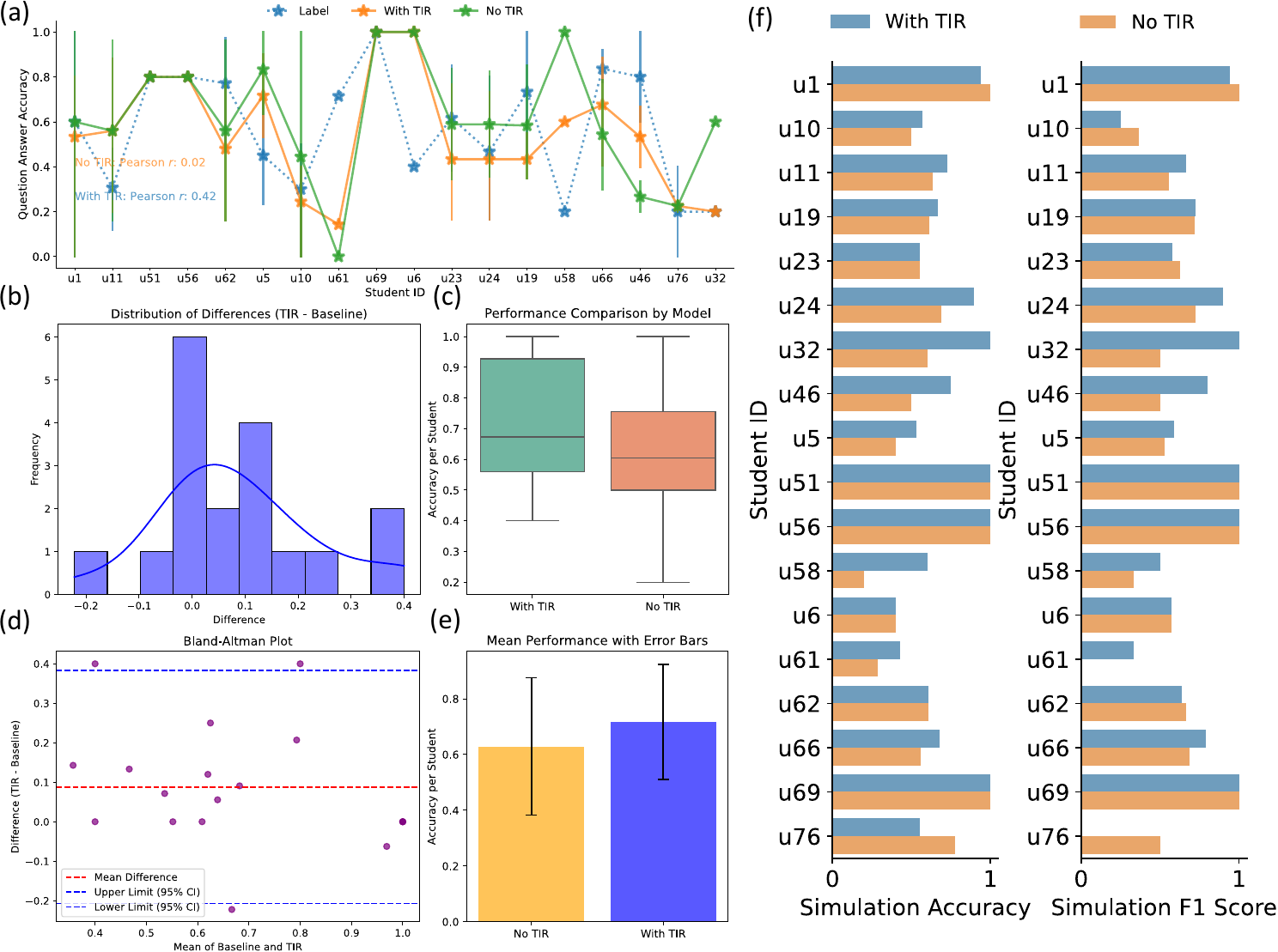}
\caption{
\mytextcolor{
Individual-level results: simulations using BertKT with and without TIR across different students.
(a). Correlation between student ID and simulated/real post-test question answer accuracy (vertical bar: standard deviation). (b,c,d,e). The distribution of simulation accuracy differences (b), boxplot (c), Bland-Altman plot to show the mean differences (d), and barplot (e) (error bar: 95\% confidence interval) between two models. (f). Average simulation accuracy and F1 score for each individual student using two models.
}}
\Description{
This figure contains six subplots (labeled a–f) presenting the performance comparison of BertKT with and without TIR for question answer accuracy in simulations. The subplots highlight various statistical and comparative insights based on the dataset.
(a) A line plot showing "Question Answer Accuracy" for different student IDs. Three sets of data are represented: the label (ground truth), simulations with TIR (orange solid line with stars), and simulations without TIR (green solid line with stars). Each point represents the average accuracy for a student, with vertical bars indicating the standard deviation. Pearson correlation coefficients for simulations with TIR (r = 0.42) and without TIR (r = 0.02) against the labels are noted on the plot.
(b) A histogram showing the distribution of differences between TIR and baseline (no TIR). The x-axis represents the difference in accuracy, while the y-axis represents the frequency of occurrences. A smooth curve overlays the histogram, indicating the approximate probability density. It shows that the data is normally distributed.
(c) A boxplot comparing "Accuracy per Student" for simulations with TIR and without TIR. The boxes represent the interquartile range, with the median shown as a line within each box, and whiskers extending to represent the range. It shows that the TIR case has better accuracy.
(d) A Bland-Altman plot visualizing the agreement between TIR and baseline performance. The x-axis represents the mean of the two methods (baseline and TIR), while the y-axis shows the difference. Dotted lines indicate the mean difference, upper, and lower limits of agreement (95\% confidence interval). Points are scattered across the plot to show individual observations.
(e) A bar chart showing mean performance for simulations with TIR and without TIR, along with error bars representing the 95\% confidence interval. It shows that the TIR case has better accuracy.
(f) Two grouped bar plots comparing simulation accuracy (left) and F1 score (right) for each student ID. The bars are color-coded: blue for simulations with TIR and orange for those without TIR. Each student ID is labeled on the y-axis. It shows that the TIR case has better accuracy.
}
\label{individual student correlation}
\end{figure*}

We then examined whether the models could capture the individual differences and correlation among simulated and real students. Specifically, we used the BertKT with or without the TIR module \mytextcolor{(baseline)} for simulation and compared with the 
label (real students' groundtruth). 
This was quantitatively measured by the Pearson correlation between the simulated students' test accuracy sequence along with student IDs and the real students'.    
As depicted in Fig. \ref{individual student correlation}, we found that the integration of the TIR module better captured the correlation between simulated and real students regarding the learning performance sequence along with student IDs than the no TIR case and apparently improved the Pearson correlation from $r=0.02$ (No TIR) to $r=0.42$ (With TIR).
These results demonstrate that our TIR module enabled more realistic simulation by better capturing the individual differences of student learning performance.

\mytextcolor{
We further checked the statistical difference of the average simulation accuracy per student with or without the TIR module (baseline).
The normality of the differences between both was assessed using the Shapiro-Wilk test, which indicated no significant deviation from normality ($W = 0.9235, \, p = 0.1493$; df = 17). A paired t-test was then conducted to evaluate the impact of the TIR module on prediction performance. The results showed a statistically significant improvement in accuracy with the TIR module compared to the baseline ($t = 2.4139,\, p = 0.0273$; df = 17). A Bland-Altman analysis revealed a mean difference (bias) of 0.0881 (95\% CI: 0.0186 to 0.1577), with the limits of agreement ranging from -0.2069 (95\% CI: -0.5100 to 0.0962) to 0.3832 (95\% CI: 0.0800 to 0.6863). 
The Bland-Altman plot (Fig. \ref{individual student correlation}(d)) visualizes these findings, showing the mean difference as a dashed red line and the limits of agreement as dashed blue lines. The scatter of points around the mean difference is relatively consistent, suggesting that the agreement between the two models does not vary substantially across the range of predicted accuracy values. 
These results indicate a consistent positive effect of the TIR module on prediction performance, while maintaining acceptable levels of agreement with the baseline model.
}

\subsection{TIR Captures Lecture Correlation Better}

\begin{figure*}
\centering
\includegraphics[width=1\linewidth]{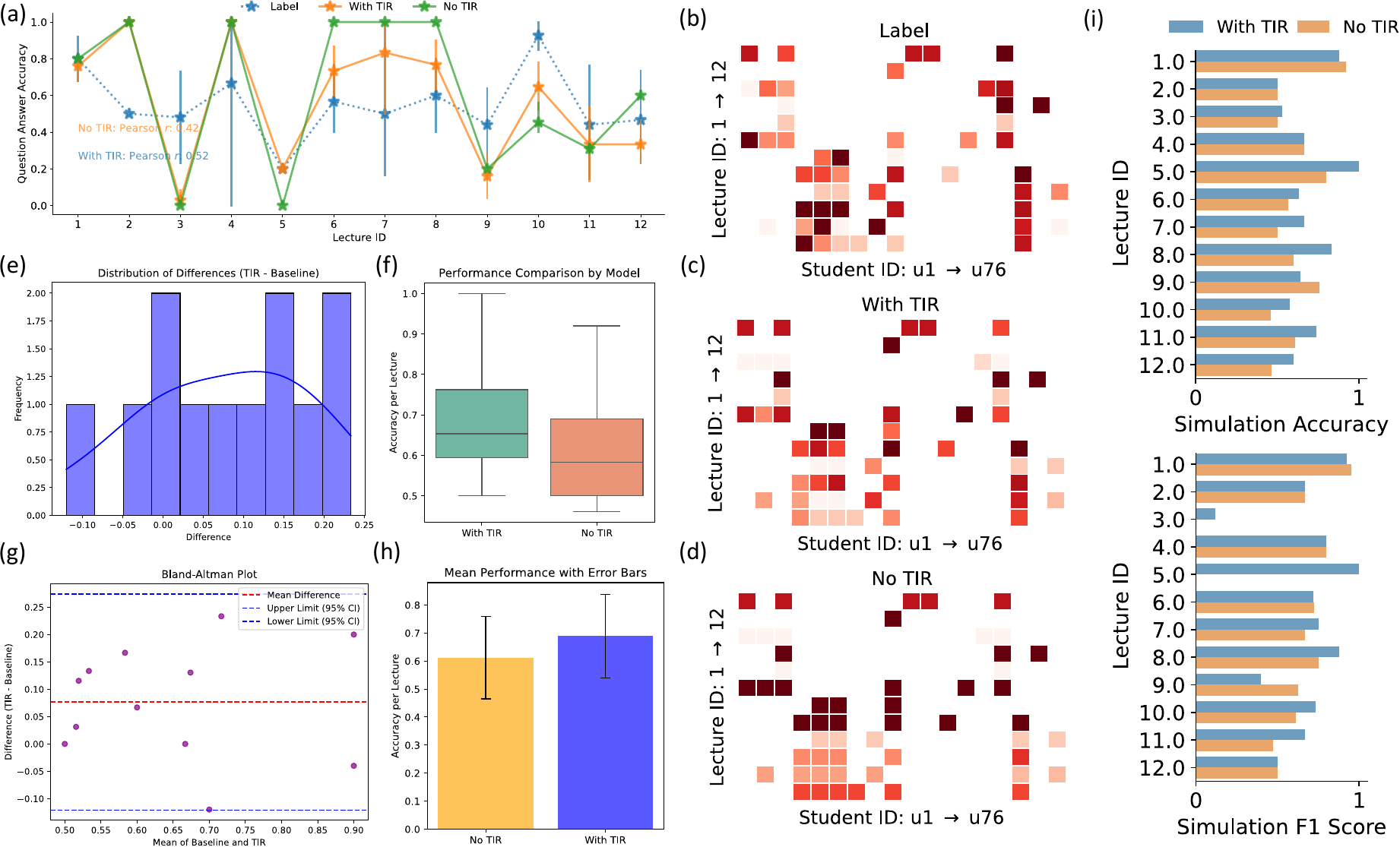}
\caption{
\mytextcolor{
Lecture-level results: simulations using BertKT with and without TIR across different lecture IDs.
(a). Correlation between lecture ID and simulated/real post-test question answer accuracy (vertical bar: standard deviation). (b)(c)(d). Heatmaps of label, BertKT with TIR, and BertKT without TIR question answer accuracy across all students and lectures. Each cell shows the average question answer accuracy of a student answering questions in a specific lecture. Darker color represents higher question answer accuracy. (e,f,g,h). The distribution of simulation accuracy differences (e), boxplot (f), Bland-Altman plot to show the mean differences (g), and barplot (h) (error bar: 95\% confidence interval) between two models. (i). Average simulation accuracy and F1 score for each lecture ID using two models.
}}
\Description{
This figure contains nine subplots (labeled a–i).
(a) shows Question Answer Accuracy Across Lectures: This line plot compares the question answer accuracy of BertKT with TIR (orange line with stars), without TIR (green line with triangles), and the label trend (dotted line) across 12 lectures. Vertical bars represent the standard deviation. The Pearson correlation coefficients are noted: 0.52 (with TIR) and 0.42 (without TIR).
(a) shows that BertKT with TIR more closely aligns with the label accuracy trend compared to the baseline (no TIR), demonstrating higher consistency. The Pearson correlation confirms a stronger association with TIR integration.
(b), (c), (d) show the Heatmaps of Accuracy per Lecture and Student:
Three heatmaps visualize question answer accuracy for the label (b), with TIR (c), and without TIR (d) across 12 lectures and all students (u1 to u76). Darker colors represent higher accuracy.
They show that TIR integration produces patterns closer to the label accuracy heatmap. Without TIR, the accuracy distribution appears more dispersed, indicating weaker alignment with ground truth.
(e) shows the Distribution of Differences (TIR - Baseline):
It shows a histogram showing the distribution of accuracy differences between TIR and baseline (no TIR). The plot includes a smooth density curve for visualizing probability. It shows that the difference is normally distributed.
(f) shows the Performance Comparison by Model (Boxplot):
It presents a boxplot comparing the accuracy per lecture for simulations with and without TIR. The median and interquartile ranges are shown, with whiskers for variability.
It shows that BertKT with TIR achieves consistently higher median accuracy and a narrower interquartile range, signifying both improved and more stable performance.
(g) is a Bland-Altman Plot (Agreement Between Methods):
This plot assesses the agreement between baseline and TIR methods. The x-axis represents the mean accuracy of both methods, and the y-axis shows their difference. Dotted lines indicate the mean difference and the limits of agreement.
It reveals that most points lie within the limits of agreement, indicating reasonable consistency between the methods. However, the positive mean difference suggests that TIR consistently outperforms the baseline.
(h) shows the Mean Performance With Error Bars:
It presents a bar chart compares the mean accuracy across all lectures for TIR and baseline methods, with error bars representing the 95\% confidence interval.
It shows that BertKT with TIR exhibits higher mean accuracy and less overlap between confidence intervals, suggesting a statistically significant improvement.
(i) shows the Simulation Accuracy and F1 Score Comparison (Grouped Bar Charts):
It presents Two bar charts compare simulation accuracy and F1 scores across 12 lectures for TIR and baseline methods.
It shows that TIR achieves higher accuracy and F1 scores in most lectures, highlighting its efficacy. 
}
\label{lecture correlation}
\end{figure*}

We then examined whether the models could capture the lecture correlation and differences. Specifically, we still used the BertKT with or without the TIR module \mytextcolor{(baseline)} for simulation and compared with the 
label (real students' groundtruth). Since different lectures had their own difficulty, students therefore had different learning performance (post-test question accuracy) across different lectures. Therefore, by comparing the trend of simulated and real students' learning performance along with the lectures, we could see whether the simulation models could capture such variations of lecture difficulty and cross-lecture correlation. 
This trend was quantitatively measured by the Pearson correlation between the simulated students'  test accuracy sequence along with lectures and the real students' sequence.    
As depicted in Fig. \ref{lecture correlation}(a), we found that the integration of the TIR module better captured the correlation between simulated and real students regarding the learning performance sequence along with lectures than the no TIR case and apparently improved the Pearson correlation from $r=0.42$ (No TIR) to $r=0.52$ (With TIR).
For more intuitive visualization in individual students, we showed the average question answering accuracy in each lecture for each specific simulated and real student, as depicted in Fig. \ref{lecture correlation}(b,c,d). This visualization also revealed larger similarity between simulated students (with TIR) and real students, compared with the simulation similarity without TIR.
These results demonstrate that our TIR module enables more realistic simulation by better capturing the lecture correlation in student learning performance.

\mytextcolor{
We further checked the statistical difference of the average simulation accuracy per lecture with or without the TIR module. The normality of the accuracy differences between them was evaluated using the Shapiro-Wilk test, which indicated no significant deviation from normality ($W = 0.9736, \, p = 0.9444$; df=11). A paired t-test was then performed to assess the impact of the TIR module on simulation performance. The analysis revealed a statistically significant improvement in accuracy with the TIR module compared to the no TIR case ($t = 2.5173, \, p = 0.0286$; df=11). Bland-Altman analysis (Fig. \ref{lecture correlation}(g)) showed a mean difference (bias) of 0.0764 (95\% CI: 0.0195 to 0.1334), with limits of agreement ranging from -0.1209 (95\% CI: -0.3263 to 0.0845) to 0.2738 (95\% CI: 0.0684 to 0.4792). 
These findings suggest that the TIR module consistently enhances prediction performance while demonstrating acceptable agreement with the baseline model.
}

\subsection{TIR Captures Question Differences Better}
\begin{figure*}
\centering
\includegraphics[width=1\linewidth]{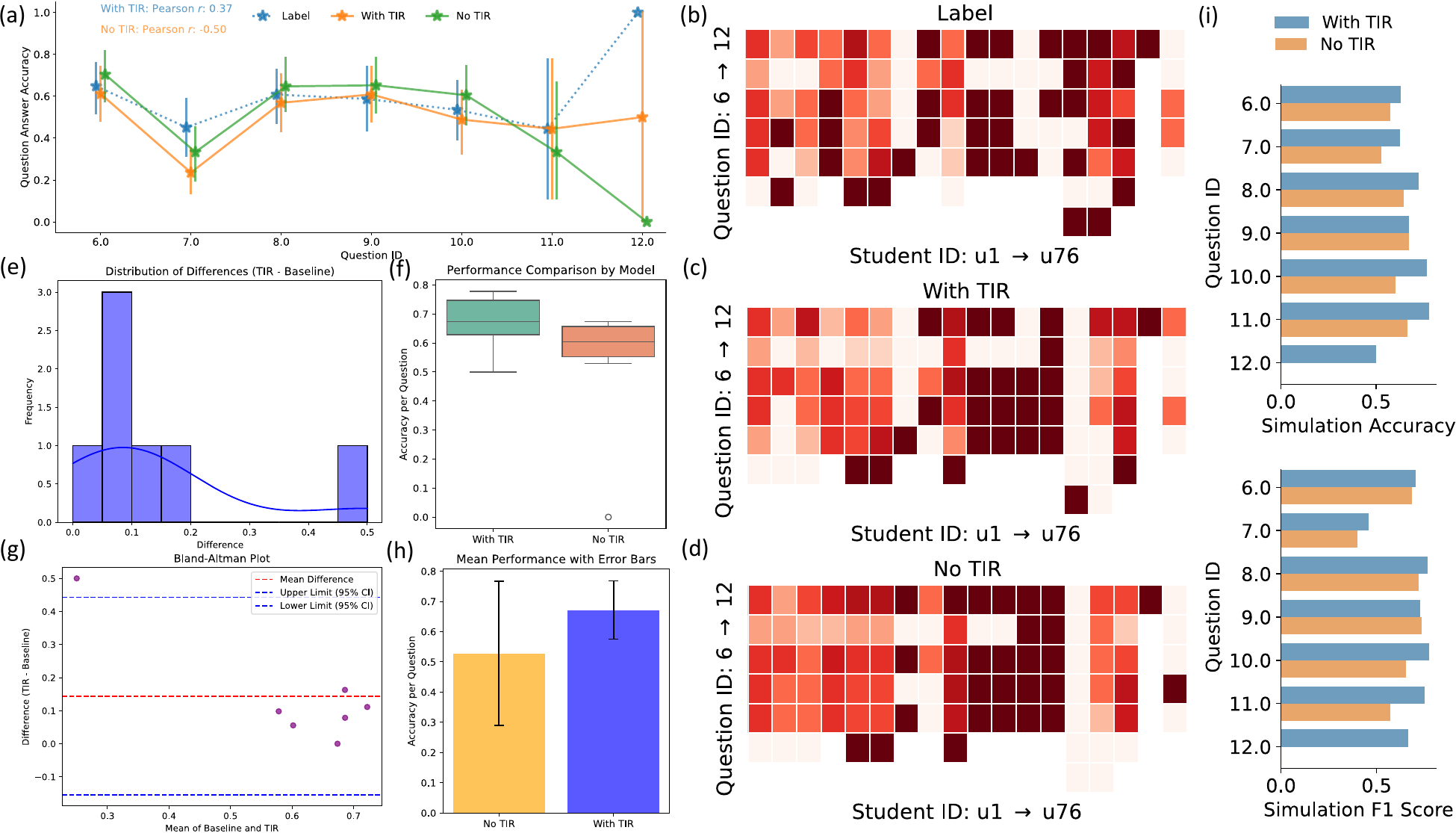}
\caption{
\mytextcolor{
Question-level results: simulations using BertKT with and without TIR across different post-test question IDs.
(a). Correlation between question ID and simulated/real post-test question answer accuracy (vertical bar: standard deviation). (b)(c)(d). Heatmaps of label, BertKT with TIR, and BertKT without TIR question answer accuracy across all students and questions. Each cell shows the average question answer accuracy of a student answering a specific question. Darker color represents higher question answer accuracy. (e,f,g,h). The distribution of simulation accuracy differences (e), boxplot (f), Bland-Altman plot to show the mean differences (g), and barplot (h) (error bar: 95\% confidence interval) between two models. (i). Average simulation accuracy and F1 score for each post-test question ID using two models.
}}
\Description{
This figure contains nine subplots (labeled a–i).
(a) shows Question Answer Accuracy Across Question IDs: This line plot compares the question answer accuracy of BertKT with TIR (orange line with stars), without TIR (green line with triangles), and the label trend (dotted line) across 7 question IDs. Vertical bars represent the standard deviation. The Pearson correlation coefficients are noted: 0.37 (with TIR) and -0.50 (without TIR).
(a) shows that BertKT with TIR more closely aligns with the label accuracy trend compared to the baseline (no TIR), demonstrating higher consistency. The Pearson correlation confirms a stronger association with TIR integration.
(b), (c), (d) show the Heatmaps of Accuracy per Question ID and Student:
Three heatmaps visualize question answer accuracy for the label (b), with TIR (c), and without TIR (d) across 7 questions and all students (u1 to u76). Darker colors represent higher accuracy.
They show that TIR integration produces patterns closer to the label accuracy heatmap. Without TIR, the accuracy distribution appears more dispersed, indicating weaker alignment with ground truth.
(e) shows the Distribution of Differences (TIR - Baseline):
It shows a histogram showing the distribution of accuracy differences between TIR and baseline (no TIR). The plot includes a smooth density curve for visualizing probability. It shows that the difference is not normally distributed.
(f) shows the Performance Comparison by Model (Boxplot):
It presents a boxplot comparing the accuracy per question ID for simulations with and without TIR. The median and interquartile ranges are shown, with whiskers for variability.
It shows that BertKT with TIR achieves consistently higher median accuracy and a narrower interquartile range, signifying both improved and more stable performance.
(g) is a Bland-Altman Plot (Agreement Between Methods):
This plot assesses the agreement between baseline and TIR methods. The x-axis represents the mean accuracy of both methods, and the y-axis shows their difference. Dotted lines indicate the mean difference and the limits of agreement.
It reveals that most points lie within the limits of agreement, indicating reasonable consistency between the methods. However, the positive mean difference suggests that TIR consistently outperforms the baseline.
(h) shows the Mean Performance With Error Bars:
It presents a bar chart compares the mean accuracy across all question IDs for TIR and baseline methods, with error bars representing the 95\% confidence interval.
It shows that BertKT with TIR exhibits higher mean accuracy and less overlap between confidence intervals, suggesting a statistically significant improvement.
(i) shows the Simulation Accuracy and F1 Score Comparison (Grouped Bar Charts):
It presents Two bar charts compare simulation accuracy and F1 scores across 7 question IDs for TIR and baseline methods.
It shows that TIR achieves higher accuracy and F1 scores in most question IDs, highlighting its efficacy. 
}
\label{question correlation}
\end{figure*}

Moreover, we further explored whether the models could capture the different questions' correlation using the BertKT with or without the TIR module. Different questions corresponded to specific knowledge concepts of course materials. Therefore, by comparing the trend of simulated and real students' test accuracy along with different questions, we could see whether the simulation models could capture students' learning performance across fine-grained and varying knowledge concepts. 
This trend was also quantitatively measured by the Pearson correlation between the simulated student question accuracy sequence along with question ID and the real students'.    
As depicted in Fig. \ref{question correlation}(a), we found that the integration of the TIR module better captured the correlation with question ID compared with real cases (label) than the no TIR case and apparently improved the Pearson correlation from $r=-0.50$ (No TIR) to $r=0.37$ (With TIR).
For more intuitive visualization in individual students, we showed the average question answering accuracy in each question for each specific simulated and real student, as depicted in Fig. \ref{question correlation}(b,c,d). This visualization also revealed larger similarity between simulated students (with TIR) and real students, compared with the simulation similarity without TIR.
These results demonstrate that our TIR module enables more realistic simulation by better capturing the question correlation (i.e. knowledge concept correlation) in student learning performance.

\mytextcolor{
We then checked the statistical difference of the average simulation accuracy per question with or without the TIR module (baseline).
The normality of the differences between both was assessed using the Shapiro-Wilk test, which indicated a significant deviation from normality ($W = 0.7451,\, p = 0.0113$; df = 6). Therefore, a Wilcoxon signed-rank test (instead of a paired t-test) was performed to evaluate the impact of the TIR module on prediction performance. The test revealed a statistically significant improvement in accuracy with the TIR module compared to the baseline ($p = 0.0277$; df = 6). A Bland-Altman analysis (Fig. \ref{question correlation}(g)) showed a mean difference (bias) of 0.1437 (95\% CI: 0.0306 to 0.2568), with the limits of agreement ranging from -0.1555 (95\% CI: -0.4754 to 0.1644) to 0.4429 (95\% CI: 0.1230 to 0.7628). These results demonstrate a significant positive effect of the TIR module on prediction performance, with an acceptable level of agreement between the two models.
}

\subsection{Dynamism of Skill Levels in Learning Path}
\label{subsec: dynamism}
\begin{figure*}
\centering
\includegraphics[width=1\linewidth]{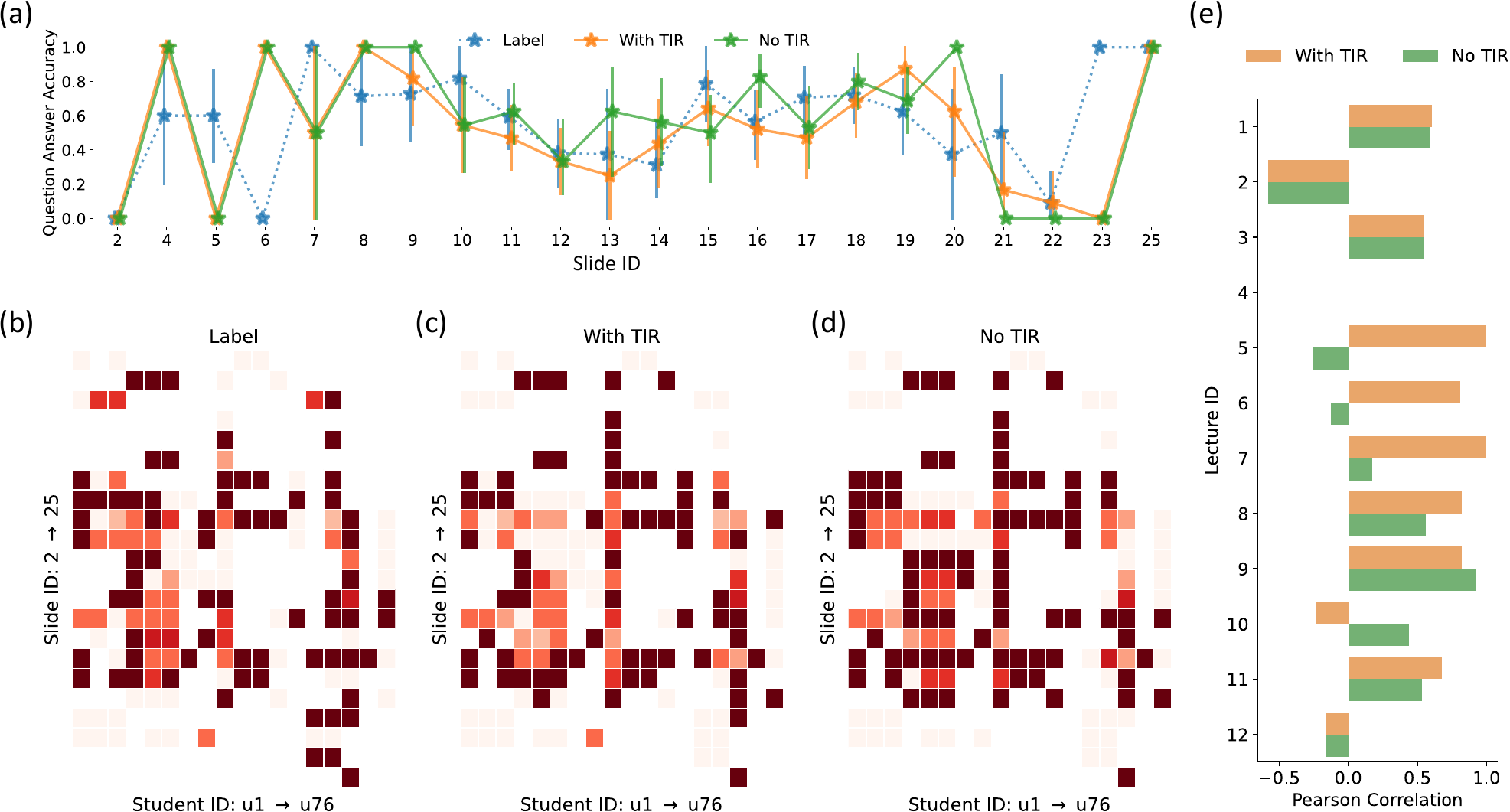}
\caption{\mytextcolor{(a). Question answer accuracy of simulations using BertKT with and without TIR against the labels on our newly collected dataset across questions related to each slide. Each star point depicts the average question answer accuracy of questions related to that slide while the vertical bar represents the standard deviation. The dotted line represents the label question answer accuracy trend. The solid lines represent the predicted question answer accuracy trend. (b)(c)(d). Heatmaps of label, BertKT with TIR, and BertKT without TIR question answer accuracy across all students and slides. Each cell shows the average student's answer accuracy of questions that are related to the slide. Darker color represents larger question answer accuracy. (e). Pearson correlation between the simulated (BertKT with or without TIR) skill level sequence along with slide ID and the real student sequence along with slide ID in each lecture on our newly collected dataset.}}
\Description{Figure (a) shows the question answer accuracy of simulations using BertKT with and without TIR against the labels on our newly collected dataset across questions related to each slide. It shows that the integration of the TIR module better captured the dynamism of skill levels in the whole learning path from the first slide to the last slide compared with real cases (label) than the no TIR case. Figure (b,c,d) show heatmaps of label, BertKT with TIR, and BertKT without TIR question answer accuracy across all students and slides. Each cell shows the average question answer accuracy related to a specific slide of a student. They show larger similarity between simulated students (with TIR) and real students, compared with the simulation similarity without TIR. Figure (e) shows the Pearson correlation between the simulated (BertKT with or without TIR) skill level sequence along with slide ID and the real student sequence along with slide ID in each lecture on our newly collected dataset. It shows that the integration of our TIR module captured better such correlation in most lectures.}
\label{slide course correlation}
\end{figure*}

Furthermore, we explored whether the simulation captured the dynamism of students' skill levels in the learning path. Here the learning path referred to the chronological learning process from the first slide to the last slide in the lecture. 
In our online education system, students' skill levels were represented by the average question answering accuracy where the questions were corresponding to a specific slide. This enabled us to measure to what extent the students mastered the knowledge concepts per slide.         
We still used the BertKT with or without the TIR module for simulation and compared with the label. Then we compared the trend of simulated and real students' skill levels along with the slide ID. 
As depicted in Fig. \ref{slide course correlation}(a), we found that the integration of the TIR module better captured the dynamism of skill levels in the whole learning path from the first slide to the last slide compared with real cases (label) than the no TIR case.
For more intuitive visualization in individual students, we also showed the average question answering accuracy in each slide for each specific simulated and real student, as depicted in Fig. \ref{slide course correlation}(b,c,d). This visualization also revealed larger similarity between simulated students (with TIR) and real students, compared with the simulation similarity without TIR.
This trend was also quantitatively measured by the Pearson correlation between the simulated skill level sequence along with slide ID and the real student sequence along with slide ID. However, since different lectures had different slides, we analyzed the Pearson correlation in each lecture. As depicted in Fig. \ref{slide course correlation}(e), the integration of our TIR module captured better correlation in most lectures.
These results demonstrate that our TIR module enables more realistic simulation by better capturing the dynamism of students' skill levels across the learning path.

\subsection{Fine-Grained Inter-Student Correlation}
\begin{figure*}
\centering
\includegraphics[width=1\linewidth]{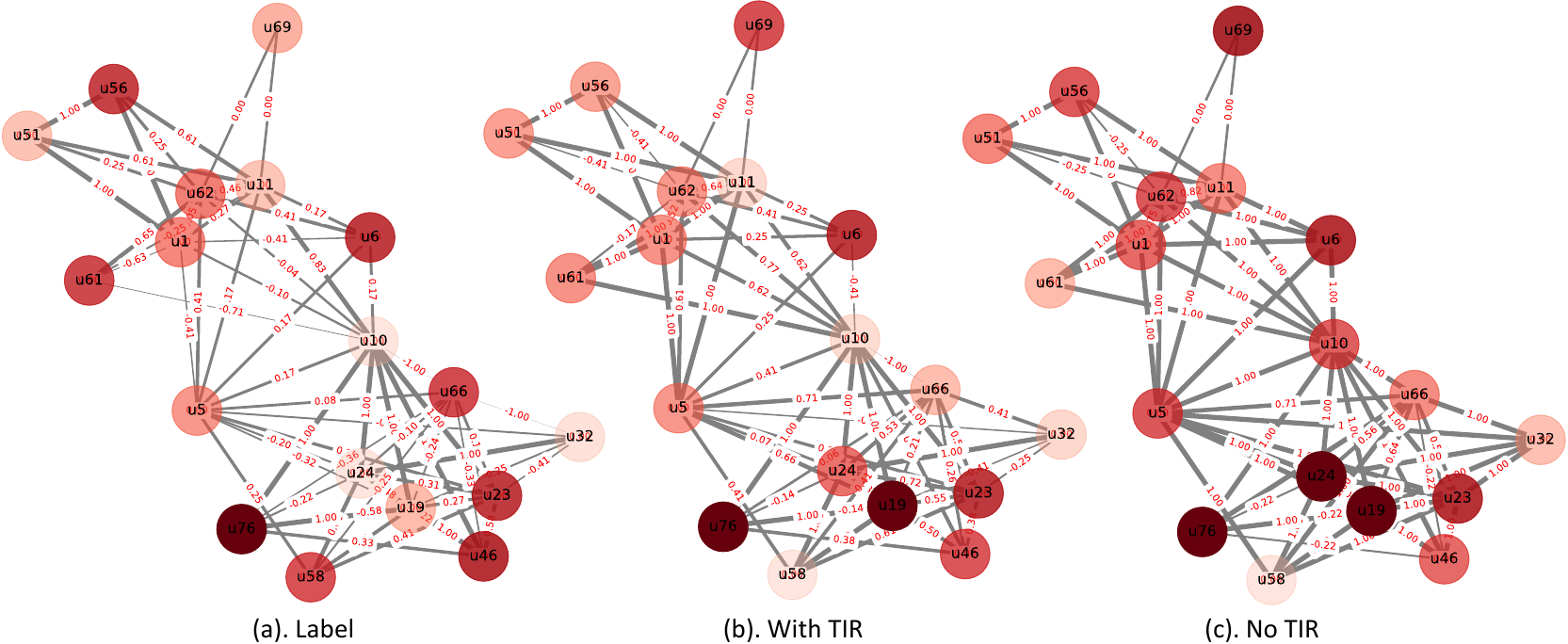}
\caption{Inter-student correlation graphs of the label (a) and simulations using BertKT with (b) and without TIR (c) on our newly collected dataset where a node is a student and an edge connects two students if both students took the same lecture. The node color depth represents the average question answering correctness of all lectures that the individual student attended. The edge weight connected by two nodes represents the inter-student correlation, which was calculated by the Pearson correlation between the question correctness sequences of two students corresponding to the questions from the overlapped lectures between two students.}
\Description{This figure shows the inter-student correlation graphs of the label (a) and simulations using BertKT with (b) and without TIR (c) on our newly collected dataset where a node is a student and an edge connects two students if both students took the same lecture. The node color depth represents the average question answering correctness of all lectures that the individual student attended. The edge weight connected by two nodes represents the inter-student correlation, which was calculated by the Pearson correlation between the question correctness sequences of two students corresponding to the questions from the overlapped lectures between two students. It shows that the integration of the TIR module better captured both individual student learning performance (average question answering correctness represented by the color depth of each node) and inter-student correlation (Pearson correlation of question correctness sequences between two students, represented by the edge weight between two nodes), which were more similar with real students (label), compared with the model without the TIR module.
}
\label{graph}
\end{figure*}

One important aspect of contextual simulation was to not only capture the individual differences in learning but also the individual correlations in the same course. Fig. \ref{graph} showed the inter-student correlation for both simulated students (BertKT with or without TIR) and real students. Each node represented one student and two nodes were connected if both students took the same lecture. The color depth of each node represented the average question answering correctness of all lectures that the individual student attended. The weight of the edge connected by two nodes represented the inter-student correlation, which was calculated by the Pearson correlation between the question correctness sequences of two students corresponding to the questions from the overlapped lectures between two students. Note that each student attended multiple lectures. But two students might not attend the same lectures. Therefore, the edge weight only considered the overlapped lectures between two students. However, the color depth of each node considered all lectures that one student had attended. That was why the edge weight might be 1 but the two nodes had different color depth. This meant that the students had the same correctness sequence for the overlapped lectures but their overall accuracy for all lectures attended by each student was different.
As depicted in Fig. \ref{graph}, we found that the integration of the TIR module better captured both individual student learning performance (average question answering correctness represented by the color depth of each node) and inter-student correlation (Pearson correlation of question correctness sequences between two students, represented by the edge weight between two nodes), which were more similar with real students (label), compared with the model without the TIR module.
These results demonstrate that our TIR module enables more realistic and finer-grained simulation by better capturing the inter-student correlation in student learning performance.

\section{Discussion}
\label{sec: discussion}

In this work, we run a 6-week online education workshop to collect fine-grained annotations of both course materials and student learning performance. This enabled a contextual student simulation to consider the effect of course materials' modulation on student learning. We further improved the student simulation by proposing a transferable iterative reflection module that augmented both prompting-based LLMs' simulation and finetuning-based LLMs' simulation, which achieved even better performance than deep learning models. This was also verified in another public dataset.

\subsection{Application Scenarios}

With the increasing importance of AI-assisted education and intelligent tutoring systems, our work could serve as the important groundwork to support a list of applications in educational context.


\subsubsection{Student: Enhancing Self-Learning}
The classroom simulacra could create digital twins about a specific student based on the past learning histories. This digital twin further emulates the student's learning performance in the future course materials and tests. This could support the self-assessment and reflections of students to set personalized goals based on their learning pace. Specifically, students could track their learning trends and progress and see how their skills have developed. As a result, it could help students reflect on their strengths and weakness to be improved during learning. Students can also set more appropriate learning goals, milestones, and study priorities based on the simulated results so that they can be motivated to achieve clearer learning targets. Such decision making and reflection in learning are also related to the metacognitive skills of students to develop learning strategies and study habits.

\subsubsection{Instructor: Adapting Teaching Strategy}
Teachers utilizing this system will be able to analyze predictions across an entire class or cohort, allowing them to test pedagogical approaches or curriculum structures before they even deploy them on a class of real human students. Having access to such a representative digital class can allow teachers to use simulation results to tailor the learning experience, presenting students with course contents that truly match their skill levels.
With the classroom simulacra, the instructors could optimize the curriculum design by identifying the common areas of difficulty, which is achieved by analyzing simulation results across the whole class.
For instance, students who turn out to learn fast can accept accelerated course pace while students who may struggle with upcoming tests or course contents can be delivered more related course resources to build personalized learning path. This could also improve the teaching resource allocation by enabling instructors to allocate teaching resources more effectively by identifying skills that may require more time for students to master. Last but not least, this system could provide insights for adaptive interventions of instructors so that students could receive personalized interventions (like verbal reminder or one-on-one tutoring) when they are identified to be at risk of falling behind.

\subsubsection{Parents: Home Support}
In the home environment, the classroom simulacra can simulate specific children's learning performance based on their past learning history. As a result, parents can have insightful information about how well their children will do in certain curriculum (strengths and weakness), and they could make better decision-making for tasks like choosing the right school, extracurricular activities, choosing advanced courses, and wisely investing in a tutoring service that can target specific areas where their children may struggle. Parents can also support learning outside of school through informed insights about necessary home activities and resources which align with their children's predicted needs and create a conducive study environment tailored to their children’s learning style according to the  performance simulation insights.

\subsection{Limitations and Future Work}
We also acknowledge the potential limitations in this work.

\subsubsection{Population Diversity}
The first limitation is the population diversity in our online education workshop. We decide to recruit students from elementary schools and high schools because these students usually do not have prior knowledge about our course materials related to Artificial Intelligence. As a result, we could better capture the learning performance and learning outcomes of students when they learn new knowledge. However, we also acknowledge that the data collection with more diverse populations (such as different age groups) could better support and further extend the findings of our experiments.


\subsubsection{Simulated Behavioral Types}
In our work, we use the students' question answering correctness to represent their learning performance, which can be mapped to students' skill levels on related course concepts. By further mapping them into specific slide IDs in the course (such as Section. \ref{subsec: dynamism}), this simulation can reflect the students' learning behaviors during the course. 
\mytextcolor{However, we acknowledge that student behaviors are not limited to question correctness. \mytextcolor{We clarify that the cognitive states information was only used for teachers to obtain students' learning states during data collection, and was not used for behavioral simulation.}
There are also more diverse learning behaviors in the real world scenarios such as reasoning processes, learning reflections, personal preferences, learning styles, etc. For example, students' cognitive states (such as attention, confusion) during the course could directly reflect their learning styles and personal preferences in the course. The simulation for such additional learning behaviors could provide further insights and evidence support. We also believe that LLMs have great potential in simulating such diverse behaviors due to the strong in-context learning ability \cite{wei2023larger} and large knowledge base \cite{alkhamissi2022review}. 
Therefore, such simulation on more diverse learning behavioral types could be the future directions for exploration. 
}

\mytextcolor{
\subsubsection{Generalization and Cost Differences}
We clarify that our goal is to show that TIR can augment LLMs to achieve better performance than themselves without TIR and deep learning models. We have a fair comparison within each LLMs-based model with or without TIR using the same training/testing data. But we do not intend to directly compare prompting-based LLMs with finetuning-based LLMs. Because both have their own pros and cons. For example, prompting-based LLMs need less training data but finetuning-based LLMs have better simulation performance. 
However, for future potential applications to extend the fine-tuned models in external datasets, it is also necessary to fine-tune models again in such new datasets, which is similar to deep learning models that use training data to update model weights. 
As such, it is not comparable/applicable to directly compare both regarding the generality or training time/computational resources. 
When comparing with deep learning, we mainly use finetuning-based LLMs for a fair comparison because they use the same training data. However, using much less training data, the TIR-augmented prompting-based LLMs can also achieve comparable or even better performance than deep learning. This also demonstrates the effectiveness of our TIR module. 
}

\mytextcolor{
\subsubsection{Insights for Educators}
Our current classroom simulacra serves as a student simulation model. Integrating it into an end-to-end intelligent teaching system entails non-trivial efforts. Nevertheless, the classroom simulacra is grounded in a real-world student-educator interaction dataset. 
\mytextcolor{Its predictive capability aligns with proven educational models that have been utilized to successfully inform teaching practices and support adaptive learning strategies \cite{scholtz2021systematic}. The foundational accuracy of our model indicates a strong potential for real-world applicability, as seen in similar simulations that have influenced educational strategies even before empirical testing \cite{xing2019dropout}}.
The simulator can support educators by delivering actionable insights that enhance personalized interventions, curriculum design, and evidence-based teaching practices. It can identify specific knowledge gaps for individual students, enabling targeted interventions, and allows educators to explore hypothetical scenarios to optimize teaching strategies for diverse learner profiles. Additionally, the simulator aids in curriculum optimization by simulating student responses to different teaching methodologies, helping to refine pacing and content sequencing. Therefore, the simulator provides a research-backed tool for testing the impact of instructional methods and predicting long-term outcomes. 
\mytextcolor{Beyond learning analytics, it integrates behavioral insights to detect learning issues and offers a safe experimental environment for innovative teaching approaches. Case studies in our work illustrate its practical utility, such as identifying impactful topics for exam preparation based on students' different learning performance on different questions (question-level) and guiding classroom time allocation based on students' learning performance across different slides (slide-level). 
In conclusion, while real-world educator experiences would strengthen our findings, the current study offers a solid foundation that demonstrates the simulator’s predictive power and practical potential. Future work that includes educator feedback will further bolster our understanding and validation of its effectiveness in real classroom settings.
}
}

\section{Conclusion}

We present Classroom Simulacra, a contextual student generative agent environment powered by large language models for learning behavioral simulation in online education. Such student agent mimics real students' past learning histories and takes actions in new course materials and test questions. This student agent environment is powered by our new fine-grained datasets and a powerful transferable iterative reflection module to augment the simulation performance. A series of experiments demonstrate the feasibility and effectiveness of our student simulation. With the increasing importance of AI-based education and intelligent tutoring systems, we believe that our work can serve as the important groundwork to support diverse applications in educational context.

\begin{acks}
This work is supported by National Science Foundation CNS-2403124, CNS-2312715, CNS-2128588 and the University of California San Diego Center for Wireless Communications.
\end{acks}

\bibliographystyle{ACM-Reference-Format}
\bibliography{sample-base}

\appendix



\section{Appendix}

\subsection{Deep Learning Baseline Models}
\label{appendix: deep learning baseline}

\begin{enumerate}
    \item Attention-Based Models: 
    \begin{itemize}
        \item AKT: The AKT model is composed of four key components: two self-attentive encoders for processing questions and knowledge acquisition, an attention-based knowledge retriever, and a feed-forward response prediction model. The question encoder generates context-aware representations of questions based on learner's past interactions, while the knowledge encoder does the same for the acquired knowledge. Then the knowledge retriever selects relevant information from the past knowledge, which is used by the response prediction model to predict learner's response to new questions\cite{10.1145/3394486.3403282}.
        \item SimpleKT: This model is similar to the AKT model. Both models fall under the attention-based category, but they differ in several aspects. SimpleKT simplifies AKT through excluding the self-attentive encoders, using dot-product function instead of time-decayed attention mechanism, and computing interaction representations without extra parameters for question difficulty. Despite these simplifications, simpleKT outperforms many deep learning models in knowledge tracing tasks\cite{DBLP:conf/iclr/0001L0H023}. 
    \end{itemize}

    \item Adversarial-Based Models:
    \begin{itemize}
        \item ATKT: The ATKT model aims to improve the generalization of knowledge tracing models by training on both original clean inputs and their adversarial examples. It consists of three key components: Interaction Projection, Attentive-LSTM, and Response Prediction. Interaction Projection maps student interactions onto interaction embeddings, providing the input for subsequent processing. The Attentive-LSTM, functioning as the backbone of the model, uses Knowledge Hidden State (KHS) modeling modules to aggregate information from prior KHSs, resulting in a composite representation. This composite representation is then fed into the Response Prediction module to generate the final predictions\cite{10.1145/3474085.3475554}.
    \end{itemize}

    \item Deep Sequential Models: 
    \begin{itemize}
        \item DKT: This model utilizes Long Short-Term Memory (LSTM) network to model learner's learning process by using a sequence of their interactions. The main idea is to represent the learner's knowledge state as a hidden state within LSTM network, capturing how the learner's knowledge evolves\cite{piech2015deepknowledgetracing}.
    \end{itemize}

    \item Memory-Augmented Models:
    \begin{itemize}
        \item DKVMN: The Dynamic Key-Value Memory Network (DKVMN) model utilizes a key-value structure to effectively track and assess a student's mastery of underlying concepts. The model consists of a static key matrix that holds the knowledge concepts and a dynamic value matrix that updates and stores the mastery levels of these concepts \cite{zhang2017dynamickeyvaluememorynetworks}.
    \end{itemize}
\end{enumerate}


\subsection{Gaze spectral clustering}
\label{appendix sub sec: gaze cluster}

The server-side aggregates fixations from all students to generate areas of interest (AoIs) feedback for the instructor. To propose these AoIs, we employ a spectral clustering method that leverages inherent geometric relationships among gaze points to analyze attended regions. We apply a Gaussian kernel to pairwise Euclidean distances to measure fixation affinity, then derive a normalized Laplacian matrix from this affinity matrix. Eigendecomposition of the Laplacian provides an eigenvalue spectrum encoding the data's underlying structure. The optimal cluster count is determined by identifying the maximal gap within the initial half of sorted eigenvalues, allowing adaptive clustering that reflects changes in the subject matter being discussed. Finally, k-means clustering is applied to the fixations to generate AoI feedback for the instructor.



\begin{figure*}
    \centering
    \includegraphics[width=1\linewidth]{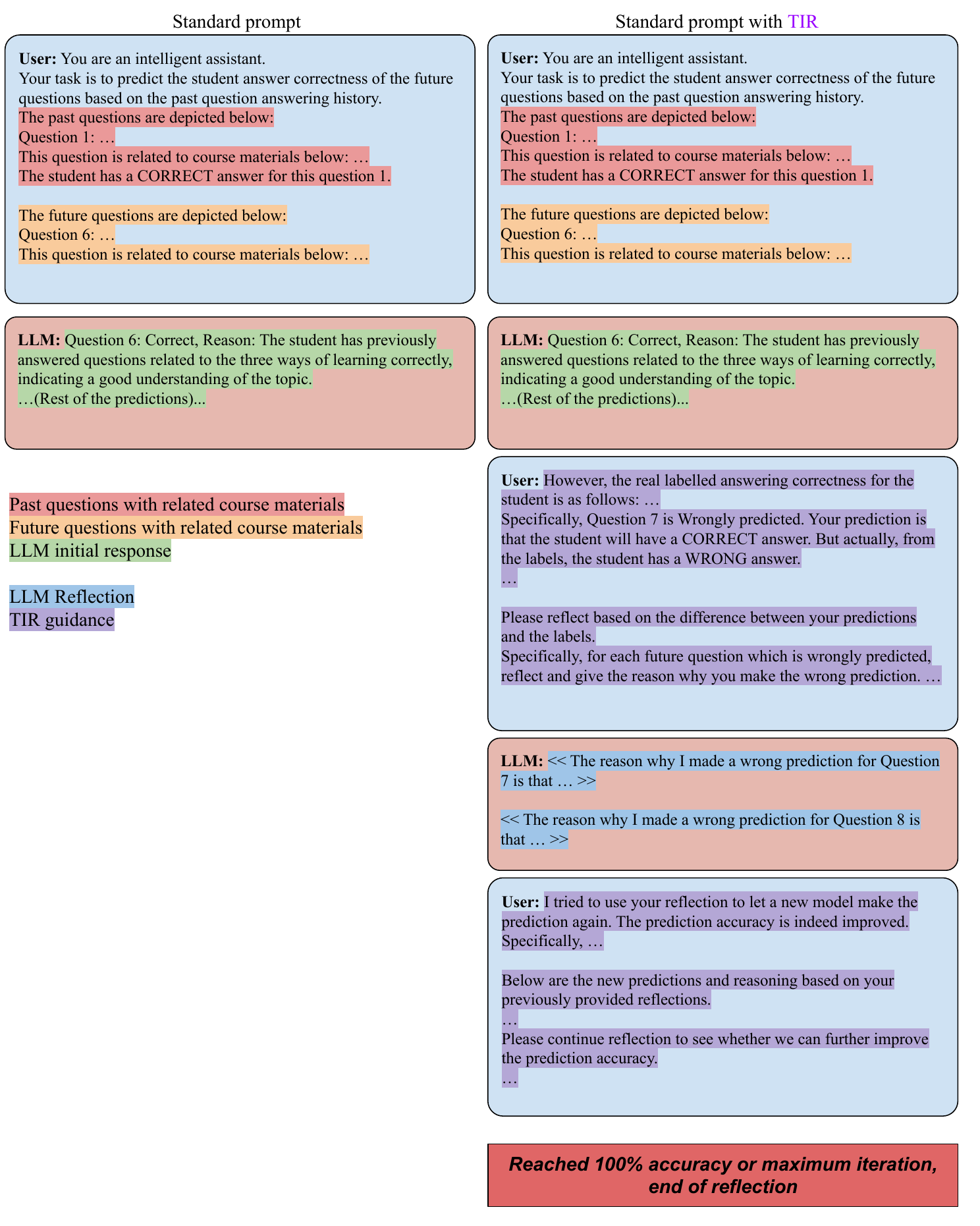}
\caption{Example prompts of standard prompting without (left) and with (right) TIR. Contents are truncated.}
    \Description{This figure shows the example prompts of standard prompting without (left) and with (right) TIR. Contents are truncated.
}
    \label{fig:std_prompt_ex}
\end{figure*}

\begin{figure*}
    \centering
    \includegraphics[width=0.75\linewidth]{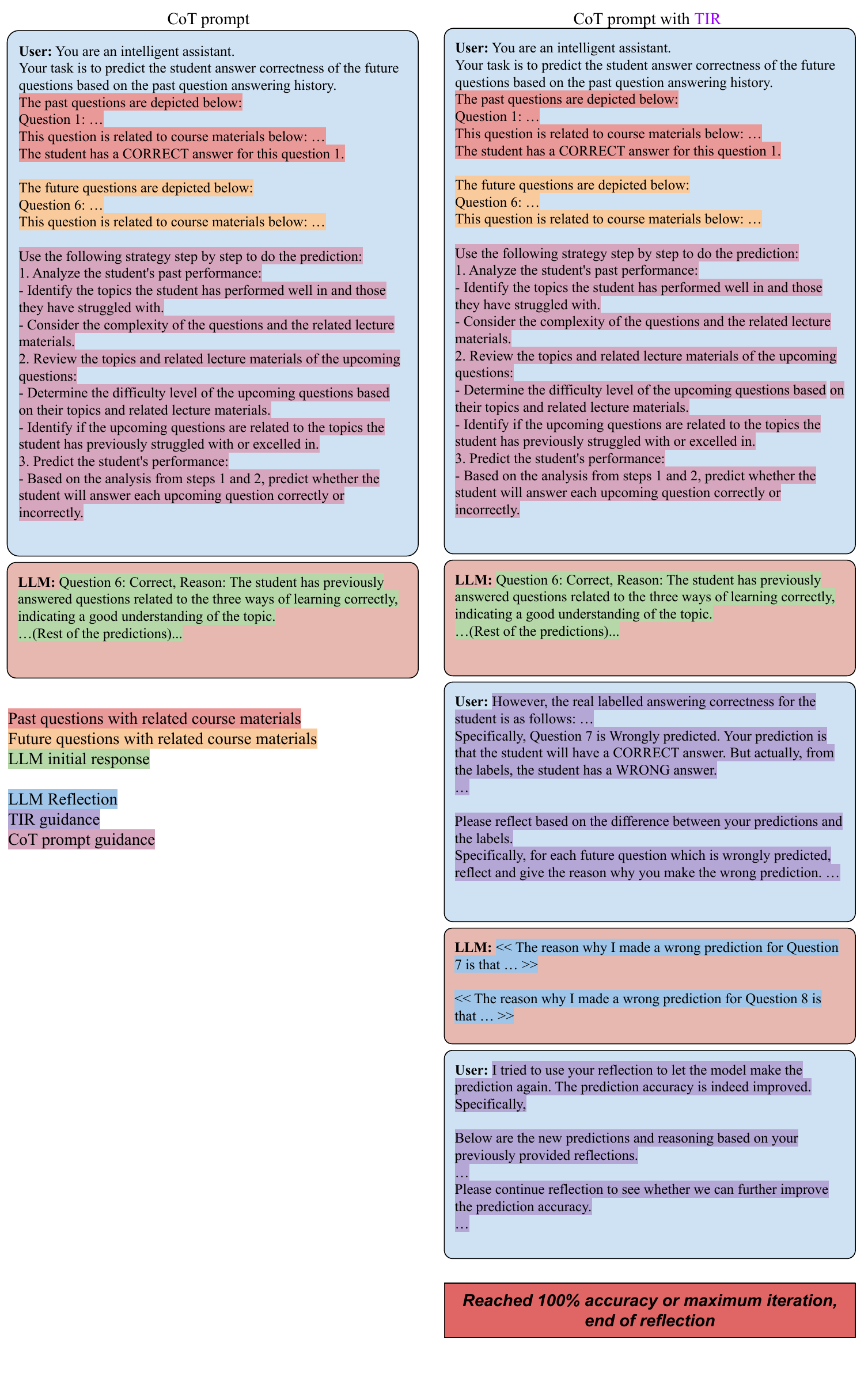}
    \caption{Example prompts of CoT prompting without (left) and with (right) TIR. Contents are truncated.}
    \Description{This figure shows the example prompts of CoT prompting without (left) and with (right) TIR. Contents are truncated.
}
    \label{fig:CoT_prompt_ex}
\end{figure*}

\begin{figure*}
    \centering
    \includegraphics[width=1\linewidth]{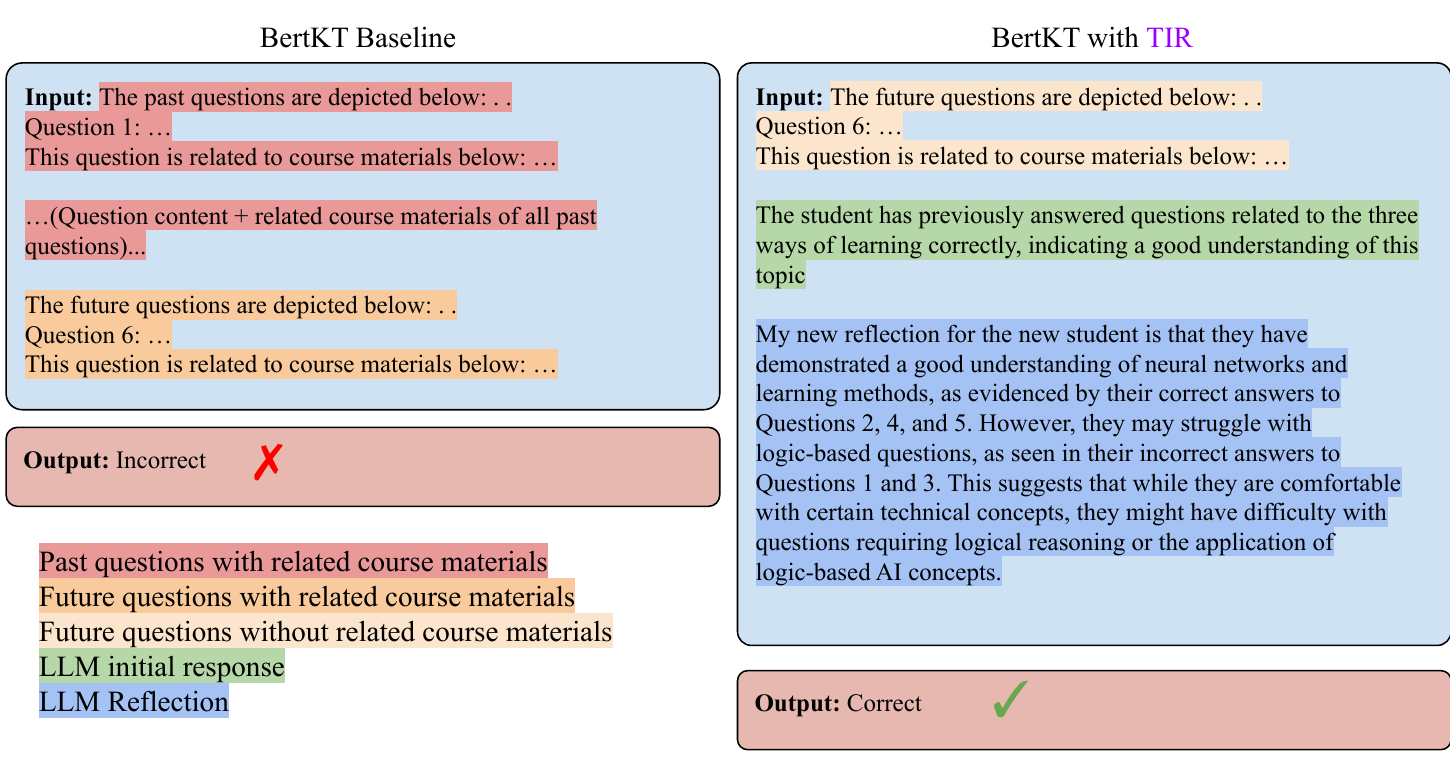}
    \caption{Example prompts of BertKT without (left) and with (right) our TIR module. The input prompts for BertKT are denoted as "Input" while the output of BertKT is denoted as "Output". Contents are truncated. Green ticks represent correct prediction while red crosses represent incorrect prediction.}
\Description{The left figure shows example prompts of BertKT without our TIR module. Its input includes text below: "
Input:
The past questions are depicted below: ...
Question 1: ...
This question is related to course materials below:...
(Question content + related course materials of all past questions)...
The future questions are depicted below:...
Question 6: ...
This question is related to course materials below:..."
The output includes text below: 
"Output: Incorrect" (which is wrong prediction according to the label)
The right figure shows example prompts of BertKT with our TIR module. Its input includes text below: "
Input: 
The future questions are depicted below: ...
Question 6: ...
This question is related to course materials below: ...
The student has previously answered questions related to the three ways of learning correctly, indicating a good understanding of this topic.
My new reflection for the new student is that they have demonstrated a good understanding of neural networks and learning methods, as evidenced by their correct answers to Questions 2, 4, and 5. However, they may struggle with logic-based questions, as seen in their incorrect answers to Questions 1 and 3. This suggests that while they are comfortable with certain technical concepts, they might have difficulty with questions requiring logical reasoning or the application of logic-based AI concepts."
The output includes text below: 
"Output: Correct" (which is right prediction according to the label).
}
\label{fig:BertKT_io_eg}
\end{figure*}

\begin{table*}[]
\caption{Workshop Syllabus}
\Description{This table shows our workshop syllabus for data collection including 12 lectures around different topics of AI.}
\label{tab:workshop syllabus}
\begin{tabular}{p{1cm} p{4cm} p{10cm}}
\hline
Lecture  & Content  & Details \\ \hline
1 & What is artificial intelligence & This is a quick introduction to AI. It inspires students with how AI builds our modern life, and also some brief ideas of how AI is capable of doing so.     \\ \hline
2 &The development of artificial intelligence&This is a further exploration of the history of AI. Students will learn about one of the most popular models used in the field: artificial neural networks (ANNs). The lecture discusses the origin and evolution of ANNs.\\ \hline
3 &Artificial neural networks that learn from teachers&This lecture presents the neural networks that learn from teachers, which is called supervised learning. Students will learn what supervised learning is and how artificial neural networks adopt this method.\\ \hline
4 &Algorithms that learn from teachers&Following the previous lecture, students dive further into the supervised learning. Many other methods also adopt this method but they are different from neural networks. Some classic algorithms will be introduced including the decision tree.\\ \hline
5 &Artificial neural networks that learn from examples (I)&This lecture introduces unsupervised learning, which means the model learns from the examples themselves, without correct answers. This lecture will show students what unsupervised learning is and an ANN model that falls into this category.\\ \hline
6 &Artificial neural networks that learn from examples (II)&This is the second part of the introduction of unsupervised learning, it contains the ANNs that make groups and compress. \\ \hline
7 &Algorithms that learn through try and error (I) &Lecture then moves to reinforcement learning. It is meant to introduce some basic methods and ideas of reinforcement learning. V-learning is introduced.\\ \hline
8 &Algorithms that learn through try and error (II) &Following the previous lecture, students are further introduced to Q-learning, another important method used in reinforcement learning. Some applications are introduced as well.\\ \hline
9 &Talking artificial neural networks&This is an introduction to voice controlled assistants. We will show students how the voice controlled assistant works.\\ \hline
10 &Seeing artificial intelligence&This is an introduction to systems that process real world images. We will show students how images are presented in computers and understood.\\ \hline
11 &Artificial and biological neural networks&This lecture serves as an introduction to the relationship between AI and biology. We will introduce the connection between AI and the human brain and even the application of brain-machine interface.\\ \hline
12 &Artificial intelligence:
Today and tomorrow&This lecture sums up the whole curriculum. It focuses on the application of AI, and also what risks it can have, so they are more interested in exploring further after the curriculum.\\ \hline
\end{tabular}
\end{table*}

\end{document}